\documentclass[manuscript,screen,nonacm]{acmart}
\usepackage[utf8]{inputenc}
\usepackage{graphicx}

\usepackage{amsmath}

\usepackage{float}
\usepackage{booktabs}
\usepackage{multirow}
\usepackage{threeparttable}
\usepackage{tabularx}
\usepackage{array}
\usepackage{enumitem}
\usepackage{xcolor}
\usepackage{textcomp}
\usepackage{tikz}
\usetikzlibrary{positioning,shapes,arrows.meta,calc,trees,fit}
\usepackage{xurl}
\newcolumntype{L}[1]{>{\raggedright\arraybackslash}p{#1}}
\newcolumntype{Y}{>{\raggedright\arraybackslash}X}

\begin{document}

\title{Combining Post-Quantum Cryptography and Quantum Key Distribution: Taxonomy,  Constructions, Proofs, Standards, Deployments and Open Problems}

\author{Mona Alotaibi}
\email{ma2315029@qu.edu.qa}
\affiliation{\institution{College of Science and Engineering, Qatar University}\city{Doha}\country{Qatar}}
\author{Noora Fetais}
\email{n.almarri@qu.edu.qa}
\affiliation{\institution{College of Science and Engineering, Qatar University}\city{Doha}\country{Qatar}}
\author{Saif Al-Kuwari}
\email{smalkuwari@hbku.edu.qa}
\affiliation{\institution{Qatar Center for Quantum Computing, College of Science and Engineering, Hamad Bin Khalifa University}\city{Doha}\country{Qatar}}

\renewcommand{\shortauthors}{Alotaibi et al.}

\begin{abstract}
Post-quantum cryptography (PQC) and quantum key distribution (QKD) address the quantum threat from opposite directions: PQC replaces vulnerable public-key algorithms with schemes believed to resist quantum cryptanalysis, while QKD derives keys whose secrecy rests on physical assumptions rather than computational ones. Because each approach has a distinct failure mode, a growing body of work combines them so that a session key remains secure if either component holds. This survey organises that body of work. We first establish the deployed baseline, hybrid classical-plus-PQC key exchange, against which QKD-augmented designs must be judged. We then propose a taxonomy of hybrid PQC--QKD designs along eight dimensions (integration layer, combiner construction, authentication scope and source, QKD topology, trusted entities, key-delivery interface, classical component, and evidence type), and use it to compare published protocols, including Muckle and its successors, MACsec and TLS integrations, and IKEv2 mechanisms standardised in RFCs 8784, 9370, and 9867. We examine key-combiner constructions in light of the KEM-combiner literature and of NIST, ETSI, and ITU-T guidance, and we map the security arguments for hybrid PQC--QKD protocols that we identified onto a common grid of security model, assumptions, permitted compromise, established guarantee, and limitations. We synthesise deployment evidence from metropolitan and national QKD networks, distinguishing the few testbeds that have run PQC in the key path from pure-QKD deployments, and we summarise the standards and policy landscape, including the positions of national security agencies that recommend against relying on QKD alone. We close with open problems: authentication of the QKD classical channel, shared-infrastructure threat models that existing proofs cover only in part, device certification versus formal proofs, formal verification of hybrid handshakes, and the lack of peer-reviewed deployment economics in the sources we reviewed.
\end{abstract}



\maketitle

\section{Introduction}\label{sec:intro}

Public-key cryptography as deployed today rests on integer factorisation and discrete logarithms, both of which Shor's algorithm solves in polynomial time on a fault-tolerant quantum computer~\cite{Shor1999-Prime-Factoring}. No such machine exists: the largest current processors have on the order of $10^2$--$10^3$ physical qubits, whereas the most recent resource estimates for breaking RSA-2048 require on the order of $10^6$ noisy physical qubits running for days~\cite{gidney2021factor,gidney2025factor}, and several thousand logical qubits for 256-bit elliptic-curve discrete logarithms~\cite{roetteler2017ecc}. Expert elicitation nevertheless places a cryptographically relevant quantum computer (CRQC) within a one-to-two-decade horizon with non-negligible probability~\cite{Mosca2018-Quantum-Timeline,mosca2022quantum}, and ``harvest now, decrypt later'' collection means that traffic with long confidentiality requirements is already exposed~\cite{nsa2021faq,Mosca2018-Quantum-Timeline}. Governments have responded with migration mandates and timelines~\cite{whitehouse2022nsm10,nsa2022cnsa2,nist2024ir8547,eu2024recommendation}.

Two technical responses exist. \emph{Post-quantum cryptography} (PQC) replaces the vulnerable hardness assumptions with problems, chiefly on structured lattices and hash functions, for which no efficient quantum algorithm is known. NIST published FIPS~203 (ML-KEM), FIPS~204 (ML-DSA), and FIPS~205 (SLH-DSA) in August 2024~\cite{NIST-FIPS203,NIST-FIPS204,NIST-FIPS205}; a fourth signature standard, FN-DSA (draft FIPS~206, from Falcon), and a second KEM, HQC (selected March 2025), remain in the standards pipeline~\cite{nist2025hqc}. PQC is software-deployable and is already used at Internet scale in hybrid classical-plus-PQC key exchange in TLS~1.3~\cite{rfc9954,rfc10024}. Its security, however, is computational and rests on assumptions whose maturity is measured in years rather than decades. \emph{Quantum key distribution} (QKD) establishes keys whose secrecy follows from quantum-mechanical principles and can be proven without computational assumptions~\cite{mayers2001unconditional,lochau1999unconditional,shorpreskill2000simple,portmann2022security}. Its guarantees, however, apply to idealised devices; real devices have been attacked through their implementation~\cite{lydersen2010hacking,xu2020secure}, QKD requires dedicated optical infrastructure with range limits of tens to a few hundred kilometres per span, it needs an authenticated classical channel that must itself be secured somehow, and several national security agencies recommend against relying on it in place of PQC~\cite{nsa2020qkd,ncsc2020quantum,agencies2024qkdposition}.

\emph{Hybrid PQC--QKD} designs combine both sources of key material so that the resulting session key remains secure as long as at least one source is uncompromised. This is the same hedging argument that motivates classical-plus-PQC hybrids~\cite{bindel2019hybrid,rfc9954}; QKD adds a third, physically grounded source. The construction is simple to state and easy to get subtly wrong: the combiner must not let a compromised or adversarially influenced input weaken the others~\cite{giacon2018kem}, the classical channel of QKD must be authenticated by something, and proofs written for idealised components do not automatically cover shared infrastructure such as trusted relay nodes or a common key-management system. A survey is warranted because this literature now spans cryptographic protocol design~\cite{dowling2020muckle,bruckner2023mucklePlus,ake2026_pqc_qkd}, network-protocol engineering~\cite{garcia2023camad,jsac2025_hybrid,Buruaga2025-VMuckle,blanco2025qkdkem}, standards from four bodies (IETF, NIST, ETSI, ITU-T), and field deployments~\cite{geitz2023hybrid,OpenQKD2024-D75,Singapore-NQSN2023}. Prior surveys cover parts of this space, and the GSMA whitepaper IG.18~\cite{gsma2024ig18} analyses hybrid QKD--PQC scenarios from a telecom-industry perspective; Section~\ref{sec:related} explains what this survey adds to that coverage.

\subsection{Scope and corpus construction}\label{sec:method}
We cover key establishment and its integration into network protocols. We do not survey QKD hardware, quantum repeaters, or PQC implementation techniques except where they bear on hybrid design; we point to dedicated surveys for those topics~\cite{xu2020secure,pirandola2020advances,cao2022evolution,ravi2024survey}.

This is an expert-curated survey, not a systematic review. We built the corpus in the steps below and report the parameters of each step so that the corpus construction is transparent; we make no claim of completeness beyond the databases, search terms, and dates stated here.

\emph{Seed set.} We started from the primary hybrid PQC--QKD protocol papers known to us (Muckle~\cite{dowling2020muckle}, Muckle+~\cite{bruckner2023mucklePlus}, VMuckle~\cite{Buruaga2025-VMuckle}, the TLS/IPsec/MACsec integrations from the Eindhoven and Madrid groups~\cite{garcia2023camad,jsac2025_hybrid,buruaga2025tls,blanco2025qkdkem,hybrid_ipsec2025}, Ferreira's AKE~\cite{ake2026_pqc_qkd}) and from the governing standards (RFC~8784/9370/9867, NIST SP~800-56C~Rev.\,2, ETSI TS~103~744, ETSI GS~QKD~004/014, ITU-T X.1714).

\emph{Snowballing.} From the seed set, we followed references backward (to the combiner, QKD-security, and network literature the designs depend on) and forward (papers citing Muckle, Muckle+, and the TLS/IPsec integrations), using Google Scholar and the IACR ePrint and arXiv listings.

\emph{Database search.} We ran the query families in Table~\ref{tab:method} on IEEE Xplore, the ACM Digital Library, SpringerLink, IACR ePrint, and arXiv. The final searches were run in August 2026. Titles and abstracts were screened against the inclusion and exclusion criteria in Table~\ref{tab:method}, and the remaining records were assessed in full text. Pure-PQC and pure-QKD work is included only where a hybrid design depends on it (combiner security, QKD implementation attacks, resource estimates).

\emph{Standards and grey literature.} We included standards and agency documents from IETF, NIST, ETSI, ITU-T, ISO/IEC, NSA, CISA, NCSC, and the 2024 joint European agencies' position, the GSMA IG.18 whitepaper, and the OpenQKD project deliverables, because they define the interfaces and constraints that the peer-reviewed designs target. Preprints are cited where no peer-reviewed version exists and are labelled as preprints in the bibliography.

\emph{Core studies and background references.} We distinguish two kinds of cited work. \emph{Core surveyed studies} meet the inclusion criteria and are classified in the taxonomy (Tables~\ref{tab:taxonomy_matrix} and~\ref{tab:taxonomy_matrix_b}), in the proof comparison (Table~\ref{tab:proofs}), or in the deployment tables (Tables~\ref{tab:testbeds} and~\ref{tab:testbeds_hybrid}). \emph{Background references} (PQC specifications, QKD security proofs and distance records, prior surveys, agency statements, and resource estimates) supply context and are not classified.

\begin{table}[t]
\caption{Query families used as a coverage check on the snowballed corpus, and the inclusion criteria applied to every candidate.}
\label{tab:method}
\small
\begin{tabularx}{\columnwidth}{@{}lY@{}}
\toprule
\textbf{Query family} & \textbf{Terms (combined with AND across families, OR within)} \\
\midrule
Hybrid & ``hybrid key exchange'', ``hybrid KEM'', ``key combiner'', ``key combination'' \\
PQC & ``post-quantum'', ``PQC'', ``ML-KEM'', ``Kyber'', ``ML-DSA'', ``Dilithium'' \\
QKD & ``quantum key distribution'', ``QKD'', ``QKDN'', ``ETSI QKD 014'' \\
Protocol & ``TLS'', ``IKEv2'', ``IPsec'', ``MACsec'', ``key management system'' \\
\midrule
\multicolumn{2}{@{}p{\dimexpr\columnwidth-2\tabcolsep}@{}}{\textbf{Inclusion}: peer-reviewed venue or recognised standards body; combines PQC and QKD, or integrates QKD/hybrid keys into a named protocol, or reports QKD-network operation with PQC in the key path.} \\
\multicolumn{2}{@{}p{\dimexpr\columnwidth-2\tabcolsep}@{}}{\textbf{Exclusion}: QKD physics without a protocol or network contribution; PQC implementation work without a hybrid or QKD context; superseded preprints.} \\
\bottomrule
\end{tabularx}
\end{table}

\subsection{Related surveys}\label{sec:related}

Table~\ref{tab:related_surveys} positions this survey against prior surveys. Bernstein and Lange~\cite{bernstein2017post} give an algorithmic overview of PQC that predates the FIPS standards and does not consider QKD. Ravi et al.~\cite{ravi2024survey} survey side-channel and fault attacks on lattice schemes; we use it as the reference for implementation security of the PQC component. On the QKD side, Xu et al.~\cite{xu2020secure} and Pirandola et al.~\cite{pirandola2020advances} review protocol security with realistic devices, Portmann and Renner~\cite{portmann2022security} review composable security, Mehic et al.~\cite{mehic2020quantum} and Cao et al.~\cite{cao2022evolution} review QKD networking (key management, routing, trusted nodes, SDN control), Dervisevic et al.~\cite{dervisevic2025quantum} give a dedicated survey of key management in QKD networks (key stores, ETSI/ITU-T key-delivery interfaces, and key-relay operation), and All\'eaume et al.~\cite{alleaume2014using} is the classic treatment of how QKD keys can be used for cryptographic purposes, including the authentication problem that hybrid designs revisit. At the protocol level, Dervisevic and Mehic~\cite{dervisevic2021ipsec} review proposals for integrating QKD keys into the IPsec architecture, but consider only classical-plus-QKD combinations and predate the hybrid RFCs. Giron et al.~\cite{giron2023mapping} give a systematic mapping of post-quantum \emph{hybrid} key exchange (classical plus PQC), and Baseri et al.~\cite{baseri2024navigating} review quantum-safe network protocols; neither covers QKD as a key source. Mehic et al.\ also review quantum cryptography in 5G networks~\cite{mehic2024fiveg}. None of these surveys treats the combination of PQC and QKD key material as its subject; the two documents that do are discussed below. Fedorov~\cite{Fedorov2023-Deploying-Hybrid} discusses the engineering case for hybrid infrastructure without protocol-level or proof-level analysis, and Ricci et al.~\cite{ricci2024hybridkeys} compare classical/PQC/QKD combinations experimentally but do not survey the field. The protocol papers by Rubio Garc\'ia et al.~\cite{jsac2025_hybrid}, Ferreira~\cite{ake2026_pqc_qkd}, and H\"ovelmanns et al.~\cite{hoevelmanns2025qkdoracles} contain substantial related-work sections (the last includes a comparison of the security analyses of earlier hybrid constructions) but are primary contributions, not surveys.

Two recent documents are closest to this survey in scope. The GSMA whitepaper IG.18~\cite{gsma2024ig18} (version~1.0, 20~October~2024, 23~pages) analyses hybrid security scenarios that combine QKD and PQC for the telecom industry. It defines the meanings of ``hybrid'' and a taxonomy of hybrid security (its Sections~2--3), describes key-combination approaches and the use of PQC or symmetric ITS methods to initialise the QKD authentication key (Section~4), summarises standardisation on key combination in ITU-T, ETSI, IETF, and NIST (Section~5), and reports four proofs of concept, including a Muckle-style hybridisation in linked European testbeds, the Toshiba hybrid cryptosystem, a hybrid TLS~1.2 implementation, and IKEv2 with RFC~8784 PPKs delivered over ETSI~014 (Section~6). IG.18 is an industry whitepaper: it does not compare formal security proofs, discuss the binding of QKD key identifiers, or record measurement conditions. Abelard et al.~\cite{abelard2026assessment} (EPJ Quantum Technology, 2026) assess provably secure hybrid AKE protocols for dual-use applications. They tabulate the Muckle family, the protocol of H\"ovelmanns et al., Ferreira's protocol, and the ETSI draft VMuckle-ES by model and security level (computational, everlasting, or information-theoretic), review QKD protocols by technology readiness and secret bits per pulse, and derive use-case recommendations for military and civilian networks; two of its authors co-authored Muckle+, Muckle\#, and VMuckle~\cite{abelard2026assessment}. This survey adds to these documents in three respects. First, it maps each formal security argument by model, assumptions, permitted compromise, established guarantee, limitations, and supporting theorem (Table~\ref{tab:proofs}), including the criticisms that the proofs have received in later work. Second, it treats the key-delivery interface and key identifiers as part of the security analysis. Third, it covers protocol-level integration in TLS, IKEv2, and MACsec, separates QKD topology, trusted entities, delivery interface, and the scope of authentication in its taxonomy, and records the evidence for each study together with the measurement conditions.

\begin{table*}[t]
\caption{Related surveys. ``cl.+PQC'' = covers classical-plus-PQC hybrids only; ``cl.+QKD'' = covers classical-plus-QKD integration only; ``Combiner'' = analyses how keys are combined; ``Proofs'' = maps security arguments; ``Std'' = covers IETF/NIST/ETSI/ITU-T documents; ``Deploy'' = synthesises operational data.}
\label{tab:related_surveys}
\footnotesize\setlength{\tabcolsep}{3pt}
\begin{tabular}{@{}llccccccl@{}}
\toprule
\textbf{Survey} & \textbf{Year} & \textbf{PQC} & \textbf{QKD} & \textbf{Hybrid} & \textbf{Combiner} & \textbf{Proofs} & \textbf{Std} & \textbf{Deploy} \\
\midrule
All\'eaume et al.~\cite{alleaume2014using}         & 2014 & -- & \checkmark & partial & -- & partial & -- & -- \\
Bernstein \& Lange~\cite{bernstein2017post}        & 2017 & \checkmark & --  & -- & -- & -- & -- & -- \\
Xu et al.~\cite{xu2020secure}                      & 2020 & -- & \checkmark & -- & -- & \checkmark & -- & -- \\
Pirandola et al.~\cite{pirandola2020advances}      & 2020 & partial & \checkmark & -- & -- & \checkmark & -- & partial \\
Mehic et al.~\cite{mehic2020quantum}               & 2020 & -- & \checkmark & -- & -- & -- & ETSI & \checkmark \\
Dervisevic \& Mehic~\cite{dervisevic2021ipsec}     & 2021 & -- & \checkmark & cl.+QKD & -- & -- & IETF, ETSI & -- \\
Cao et al.~\cite{cao2022evolution}                 & 2022 & -- & \checkmark & -- & -- & -- & ETSI, ITU-T & \checkmark \\
Portmann \& Renner~\cite{portmann2022security}     & 2022 & -- & \checkmark & -- & -- & \checkmark & -- & -- \\
Fedorov~\cite{Fedorov2023-Deploying-Hybrid}        & 2023 & partial & partial & \checkmark & -- & -- & -- & partial \\
Giron et al.~\cite{giron2023mapping}               & 2023 & \checkmark & -- & cl.+PQC & \checkmark & partial & IETF & -- \\
Baseri et al.~\cite{baseri2024navigating}          & 2024 & \checkmark & partial & cl.+PQC & -- & -- & IETF & -- \\
Ravi et al.~\cite{ravi2024survey}                  & 2024 & \checkmark & -- & -- & -- & -- & -- & -- \\
Dervisevic et al.~\cite{dervisevic2025quantum}     & 2025 & -- & \checkmark & -- & -- & -- & ETSI, ITU-T & \checkmark \\
GSMA IG.18~\cite{gsma2024ig18} (whitepaper)        & 2024 & \checkmark & \checkmark & \checkmark & partial & -- & ITU-T, ETSI, IETF, NIST & PoCs \\
Abelard et al.~\cite{abelard2026assessment}         & 2026 & \checkmark & \checkmark & \checkmark & partial & \checkmark & ETSI (draft TS) & Use cases \\
\midrule
This survey & 2026 & \checkmark & \checkmark & \checkmark & \checkmark & \checkmark & IETF, NIST, ETSI, ITU-T, ISO & \checkmark \\
\bottomrule
\end{tabular}
\end{table*}

\subsection{Contributions} This survey makes five contributions.
\begin{enumerate}[leftmargin=*]
  \item It establishes the deployed baseline, hybrid classical-plus-PQC key exchange, and frames QKD-augmented designs by what they add to it, rather than treating PQC and QKD as alternatives (Section~\ref{sec:foundations}).
  \item It proposes an eight-dimension taxonomy of hybrid PQC--QKD designs that separates QKD topology, trusted entities, and key-delivery interfaces, distinguishes the authentication of the endpoint exchange from that of the QKD and key-management channels, and allows a study to carry several evidence types; it classifies each core surveyed study in it (Section~\ref{sec:taxonomy}).
  \item It analyses key-combiner constructions against the KEM-combiner literature and standards guidance, and maps the security arguments for hybrid PQC--QKD protocols that we identified onto a common grid of security model, assumptions, permitted compromise, established guarantee, limitations, and supporting theorem (Section~\ref{sec:combiners}).
  \item It compares protocol integrations at three layers (TLS~1.3, IKEv2/IPsec, MACsec) using the actual standards that govern them (Section~\ref{sec:protocols}), and synthesises deployment evidence while distinguishing pure-QKD testbeds from those that ran PQC in the key path (Section~\ref{sec:deployments}).
  \item It summarises the standards and policy landscape, including agency positions critical of QKD (Section~\ref{sec:standards}), and derives an open-problem agenda from the preceding analysis (Section~\ref{sec:open}).
\end{enumerate}
The added value of this survey, relative to the surveys in Table~\ref{tab:related_surveys}, GSMA IG.18, and Abelard et al., is therefore not a new protocol but an evidence-based comparison: it states what each proof establishes and under which compromise, which parts of a deployed system lie outside every proof we reviewed, and which measurements support which performance claims.

\subsection{Organization} The rest of this survey is organized as follows. Section~\ref{sec:foundations} reviews the quantum threat, PQC, QKD, and the security properties that a hybrid argument can establish. Section~\ref{sec:taxonomy} defines the taxonomy and classifies the core studies. Section~\ref{sec:combiners} analyses key combiners, security proofs, and shared-infrastructure threats. Section~\ref{sec:protocols} examines integration in TLS~1.3, IKEv2/IPsec, and MACsec. Section~\ref{sec:deployments} reviews deployment evidence and the reported measurements. Section~\ref{sec:standards} summarises standards and policy, and Section~\ref{sec:deployment} discusses deployment considerations. Section~\ref{sec:open} states open problems, and Section~\ref{sec:conclusion} concludes.

\section{Foundations}\label{sec:foundations}

\subsection{The quantum threat, briefly}

Shor's algorithm breaks RSA and elliptic-curve cryptography on a CRQC~\cite{Shor1999-Prime-Factoring}. Resource estimates have fallen by roughly an order of magnitude each half-decade: Fowler et al.~\cite{fowler2012surface} discussed surface-code overheads of $10^8$--$10^9$ physical qubits; Gidney and Eker\aa~\cite{gidney2021factor} reduced RSA-2048 to about 20 million noisy qubits over eight hours; Gidney~\cite{gidney2025factor} reports under one million qubits over about a week; and Roetteler et al.~\cite{roetteler2017ecc} estimate roughly 2{,}330 logical qubits for 256-bit ECC. Grover's algorithm~\cite{grover1996fast} gives a quadratic speed-up for exhaustive key search, so AES-256 retains a comfortable margin; quantum collision finding on an $n$-bit hash costs about $2^{n/3}$ queries~\cite{bht1998collision}, so SHA-256 retains roughly 85-bit collision and 128-bit preimage security. Symmetric primitives and hash functions therefore survive with parameter upgrades, which is why the surveyed hybrid designs hand their combined key to a symmetric cipher.

Mosca's framing~\cite{Mosca2018-Quantum-Timeline} makes the migration argument precise: if data must remain confidential for $x$ years and migration takes $y$ years, a CRQC arriving within $x+y$ years is a problem today, because the ciphertext can be recorded now and decrypted later.

\subsection{Post-quantum cryptography}

NIST's process ran from 2016 to 2024. The third round had seven finalists (Kyber, NTRU, SABER, Classic McEliece, Dilithium, Falcon, Rainbow) and eight alternates (BIKE, FrodoKEM, HQC, NTRU~Prime, SIKE, GeMSS, Picnic, SPHINCS+)~\cite{nist2022round3report}; in July 2022 NIST selected Kyber, Dilithium, Falcon, and SPHINCS+ for standardisation, with BIKE, Classic McEliece, HQC, and SIKE continuing to a fourth round (SIKE was subsequently broken). FIPS~203/204/205 followed in August 2024, HQC was selected in March 2025, and FN-DSA (FIPS~206) is in draft as of mid-2026~\cite{NIST-FIPS203,NIST-FIPS204,NIST-FIPS205,nist2025hqc}.

Table~\ref{tab:pqc} summarises the standardised algorithms. ML-KEM~\cite{bos2018kyber} and ML-DSA~\cite{ducas2018dilithium} rest on Module-LWE/Module-SIS~\cite{regev2009lattices,langlois2015worst}, with small (centred-binomial or uniform) noise rather than discrete Gaussians; ML-KEM reaches IND-CCA2 security through a Fujisaki--Okamoto transform, and FIPS~204 targets strong unforgeability (SUF-CMA). SLH-DSA~\cite{bernstein2015sphincs,bernstein2019sphincsplus} relies only on hash-function properties, at the cost of large signatures. HQC~\cite{hqc_spec,sendrier2017code} offers a code-based alternative to lattice assumptions. Implementation security of these schemes is a field in itself: masking of lattice KEMs~\cite{tches2022_masking}, fault attacks on ML-DSA~\cite{tches2024_dilithium_fault}, and hardware-accelerator backdoors~\cite{acmccs2024_repqc} are surveyed by Ravi et al.~\cite{ravi2024survey}. The point relevant to hybrid design is that the PQC component in a QKD node is subject to these attacks, and that a co-located implementation flaw is one of the ways in which the ``independent failure'' assumption of hybrid arguments can break (Section~\ref{sec:threats}).

\begin{table*}[t]
\caption{NIST post-quantum algorithms. Sizes are those of the published FIPS documents; timings are omitted because they depend on platform and implementation and should be taken from a named benchmark (e.g., the eBACS/SUPERCOP or liboqs results cited in the source papers).}
\label{tab:pqc}
\footnotesize\setlength{\tabcolsep}{4pt}
\begin{tabular}{@{}llrrrrl@{}}
\toprule
\textbf{Algorithm} & \textbf{Family / role} & \textbf{Cat.} & \textbf{PK (B)} & \textbf{SK (B)} & \textbf{CT / Sig (B)} & \textbf{Status} \\
\midrule
ML-KEM-512  & Lattice KEM        & 1 &  800 & 1632 &  768 & FIPS 203 (2024) \\
ML-KEM-768  & Lattice KEM        & 3 & 1184 & 2400 & 1088 & FIPS 203 (2024) \\
ML-KEM-1024 & Lattice KEM        & 5 & 1568 & 3168 & 1568 & FIPS 203 (2024) \\
ML-DSA-44   & Lattice signature  & 2 & 1312 & 2560 & 2420 & FIPS 204 (2024) \\
ML-DSA-65   & Lattice signature  & 3 & 1952 & 4032 & 3309 & FIPS 204 (2024) \\
ML-DSA-87   & Lattice signature  & 5 & 2592 & 4896 & 4627 & FIPS 204 (2024) \\
SLH-DSA-SHA2-128s & Hash signature & 1 & 32 & 64 & 7856 & FIPS 205 (2024) \\
SLH-DSA-SHA2-128f & Hash signature & 1 & 32 & 64 & 17088 & FIPS 205 (2024) \\
SLH-DSA-SHA2-192s & Hash signature & 3 & 48 & 96 & 16224 & FIPS 205 (2024) \\
SLH-DSA-SHA2-256s & Hash signature & 5 & 64 & 128 & 29792 & FIPS 205 (2024) \\
FN-DSA-512 (Falcon)  & Lattice signature & 1 &  897 & 1281 &  666 & Draft FIPS 206 \\
FN-DSA-1024 (Falcon) & Lattice signature & 5 & 1793 & 2305 & 1280 & Draft FIPS 206 \\
HQC-128 & Code KEM & 1 & 2249 & 2305 & 4433 & Selected 2025; draft pending \\
HQC-192 & Code KEM & 3 & 4522 & 4586 & 8978 & Selected 2025; draft pending \\
HQC-256 & Code KEM & 5 & 7245 & 7317 & 14421 & Selected 2025; draft pending \\
\bottomrule
\end{tabular}
\begin{tablenotes}\small
\item Cat.\ = NIST security category (1/3/5 $\approx$ AES-128/192/256 key search; 2 $\approx$ SHA-256 collision). Sources: \cite{NIST-FIPS203,NIST-FIPS204,NIST-FIPS205,falcon_spec,hqc_spec}. Falcon sizes are from the v1.2 specification and HQC sizes from the fourth-round specification; the draft FIPS 206 and the forthcoming HQC standard may change encodings, and the table should be re-checked against them when they are published.
\end{tablenotes}
\end{table*}

The relevant comparison point for any PQC--QKD design is not classical cryptography but the classical-plus-PQC hybrid that is already deployed. TLS~1.3 hybrid key exchange concatenates an ECDH shared secret with an ML-KEM shared secret and feeds the concatenation into the TLS key schedule~\cite{rfc9954,rfc10024}; major browsers, content-delivery networks, and OpenSSL ship X25519MLKEM768 by default. IKEv2 supports additional key exchanges through the IKE\_INTERMEDIATE mechanism~\cite{rfc9242,rfc9370}. Benchmarks of PQ key exchange and authentication in TLS~\cite{paquin2020benchmarking,sikeridis2020pq} show that ML-KEM key exchange adds little latency, whereas PQ certificates and signatures dominate handshake size. QKD-augmented designs inherit this baseline and add a third key source; their cost must be measured against it.

\subsection{Quantum key distribution}

QKD lets two parties who share an authenticated classical channel agree on a key whose secrecy can be proven without computational assumptions~\cite{bennett1984quantum,ekert1991quantum,mayers2001unconditional,lochau1999unconditional,shorpreskill2000simple}; modern proofs are composable and account for finite key lengths~\cite{renner2008security,tomamichel2017largely,portmann2022security}. Table~\ref{tab:qkd} compares the protocol families that appear in deployments. Decoy states~\cite{Lo2005-Decoy} make BB84 practical with attenuated lasers; measurement-device-independent QKD~\cite{Lo2012-MDI} removes detector side channels and has been demonstrated over 404~km of fibre~\cite{yin2016mdi404}; twin-field QKD~\cite{Lucamarini2018-TF} beats the repeaterless PLOB bound~\cite{plob2017} and has reached 830~km in the field~\cite{Wang2022-TF830km} and 1{,}002~km in the laboratory~\cite{Liu2023-TF1000km}; continuous-variable QKD~\cite{Grosshans2002-CV} uses coherent detection with telecom components and has been demonstrated at 100~km with a local local oscillator~\cite{Hajomer2024-CV-100km} and in the field~\cite{williams2024field}; and device-independent QKD has been demonstrated in the laboratory~\cite{nadlinger2022di}. Secure key rates above 100~Mb/s have been shown at short range~\cite{Li2023-High-Rate}, and satellite links have been demonstrated~\cite{liao2017satellite,chen2021integrated}.

\begin{table*}[t]
\caption{QKD protocol families relevant to hybrid deployment. Distances and rates are record or representative values from the cited papers, not operational guarantees.}
\label{tab:qkd}
\footnotesize\setlength{\tabcolsep}{4pt}
\begin{tabular}{@{}L{2.0cm}L{2.5cm}L{2.1cm}L{2.1cm}L{2.7cm}L{2.1cm}@{}}
\toprule
\textbf{Family} & \textbf{Physical basis} & \textbf{Demonstrated reach} & \textbf{Security model} & \textbf{Implementation-security status} & \textbf{Deployment maturity} \\
\midrule
Decoy-state BB84~\cite{bennett1984quantum,Lo2005-Decoy} & Single-photon-level polarisation/phase encoding, attenuated laser & Metro and $\sim$100--200~km spans; longer with trusted relays & Composable, finite-key~\cite{tomamichel2017largely} & Detector attacks known and mitigated case by case~\cite{lydersen2010hacking,xu2020secure}; protection profile ETSI GS QKD 016~\cite{etsi_qkd_016} & Commercial; basis of most testbeds (Table~\ref{tab:testbeds}) \\
Entanglement-based (E91/BBM92)~\cite{ekert1991quantum,BBM92} & Entangled photon pairs & Metro; satellite-to-ground~\cite{liao2017satellite} & Composable & Source-independent variants; detector side channels remain & Research and satellite demos \\
MDI-QKD~\cite{Lo2012-MDI} & Bell-state measurement at untrusted relay & 404~km lab~\cite{yin2016mdi404} & Composable; immune to detector attacks & Removes the detector attack surface; source-side assumptions remain & Field trials \\
Twin-field QKD~\cite{Lucamarini2018-TF} & Single-photon interference at untrusted relay & 830~km field~\cite{Wang2022-TF830km}, 1{,}002~km lab~\cite{Liu2023-TF1000km} & Composable; beats PLOB bound~\cite{plob2017} & Phase stabilisation; source assumptions & Field trials \\
CV-QKD~\cite{Grosshans2002-CV} & Coherent-state quadratures, homodyne/heterodyne detection & 100~km lab~\cite{Hajomer2024-CV-100km}; metro field trials~\cite{williams2024field} & Composable finite-size proofs maturing & Telecom components; excess-noise sensitivity & Field trials; some commercial products \\
DI-QKD~\cite{nadlinger2022di} & Loophole-free Bell test & Metres (lab) & Device-independent & Removes device modelling & Laboratory only \\
\bottomrule
\end{tabular}
\end{table*}

Three properties of QKD shape every hybrid design and are often understated in hybrid papers. First, \emph{QKD needs authentication}. The classical reconciliation channel must be authenticated; if it is authenticated with pre-shared keys and information-theoretic MACs, the system needs an out-of-band bootstrapping step, and if it is authenticated with public-key signatures, the information-theoretic guarantee holds only for keys established before those signatures are broken~\cite{alleaume2014using}. Post-quantum signatures are the natural way to bootstrap and to secure the key-management plane, and this is exactly what several ``hybrid'' testbeds actually did~\cite{yang2021pqcauth,geitz2023hybrid}. Second, \emph{proofs are about models, not devices}. Detector blinding~\cite{lydersen2010hacking}, source imperfections~\cite{practical_sajeed_2015}, and other implementation attacks~\cite{xu2020secure} are why certification frameworks (ETSI GS QKD 016~\cite{etsi_qkd_016}, ISO/IEC 23837~\cite{iso23837}) exist and why agencies treat QKD as unproven at the product level~\cite{nsa2020qkd,ncsc2020quantum,agencies2024qkdposition}. Third, \emph{range is extended by trusted relays}. Each multi-span metropolitan or national network in Table~\ref{tab:testbeds} relays keys through nodes that hold them in the clear; security then rests on the physical and procedural protection of those nodes~\cite{mehic2020quantum,cao2022evolution}. MDI- and TF-QKD remove trust in the relay's \emph{measurement} but not the need for a relay chain over long distances.

\subsection{Security properties and arguments}\label{sec:secprops}

Hybrid papers often make assertions about the security of a protocol along the lines of ``secure if either component is secure,'' without clarifying which security property they mean. We identify three security properties because the surveyed proofs establish them under different assumptions. 

\emph{Key secrecy} (key indistinguishability) means that session keys output by a protocol are indistinguishable from random strings to an attacker who does not trivially compromise the session. Key secrecy is a key concept in game-based security models for AKE, as shown in Table~\ref{tab:proofs}~\cite{canetti2001analysis,krawczyk2005hmqv,dowling2020muckle}. Statements of key secrecy include assumptions or restrictions on other protocol attacks that the attacker could perform, such as obtaining other session keys, long-term keys, session information, or QKD keys~\cite{hoevelmanns2025qkdoracles}. 

\emph{Peer authentication} is the property of a system that ensures one party's participation in only sessions with its intended peer; game-based definitions of peer authentication include matching sessions and no malicious acceptance. Whether a given system has peer authentication depends on the type of authentication, such as pre-shared keys or signatures or KEM-based authentication, and is separate from key-exchange security; a given system might be key-exchange secure against key compromise from a subsequent PQ attack but not have peer authentication against an adversary who can forge PQ signatures during key exchange. 

\emph{Application-data confidentiality} is the property that application data encrypted with the session key remain secret. The integrations described in Section~\ref{sec:protocols} use a computationally secure AEAD scheme, AES-GCM or ChaCha20-Poly1305, to encrypt application data. Even if the shared key is information-theoretically secret, the application data are only computationally secret. Correct use of the AEAD scheme is required for application-data confidentiality; the key-exchange proofs do not prove this property. 

The two sources are secure in different ways. A KEM is $(t,\varepsilon)$-IND-CCA secure if no adversary running in time $t$ can guess the key encapsulated by the KEM with advantage $\varepsilon$ over random guessing, given a hard assumption. A QKD protocol is $\varepsilon_{\mathrm{QKD}}$-secure if there is a probability distribution on the keys such that the trace distance between the key and an ideal key not correlated with the adversary's quantum information is at most $\varepsilon_{\mathrm{QKD}}$~\cite{renner2008security,portmann2022security}, with no computational assumptions about the adversary. 

Hybrid systems we review are analyzed with case analyzes rather than measures of component failure risk. Security analyzes of hybrid systems provide security guarantees that include limits on the adversarial advantage against which the PQC implementation is secure and limits on the adversarial advantage against which the QKD implementation is secure. Claimed security guarantees for hybrid systems differ by hybrid system and analysis approach. Muckle's main theorem claims a limit on the hybrid advantage equal to the sum of the PQC and QKD limits; H\"ovelmanns et al.\ suggest that a limit on the hybrid advantage equal to the lesser of the PQC and QKD limits could be claimed~\cite{dowling2020muckle,hoevelmanns2025qkdoracles}. H\"ovelmanns et al.\ develop game-theoretic models for both cases and show an information-theoretic limit for the QKD case and a computational limit for the PQC case under the quantum random oracle model~\cite{hoevelmanns2025qkdoracles}. Clermont and Henrich claim a limit on the hybrid advantage equal to twice the lesser of the PQC and QKD limits (Theorem~1) and show information-theoretic key secrecy under the random oracle model when the QKD implementation does not fail (Theorem~2)~\cite{clermont2026best}. We provide a table of claimed hybrid security guarantees and related assumptions and theorems for other reviewed hybrid systems in Table~\ref{tab:proofs} rather than a summary security guarantee. All reviewed analyzes assume that assumptions of component security models are met, including the absence of information leakage by the KEM implementation beyond that modeled, correct behavior of QKD and key-delivery interfaces as modeled (multiple analyzes assume perfectly uniform honest QKD keys and ignore $\varepsilon_{\mathrm{QKD}}$~\cite{hoevelmanns2025qkdoracles}), and any adversarial compromises that occur would be captured by the cleanness predicate in the model. Other infrastructure-sharing risks are discussed in Section~\ref{sec:threats}.

\section{A Taxonomy of Hybrid PQC--QKD Designs}\label{sec:taxonomy}

\begin{figure*}[t]
\centering
\resizebox{\textwidth}{!}{%
\begin{tikzpicture}[
  font=\footnotesize,
  dim/.style={draw, rounded corners, fill=black!5, minimum width=2.6cm, minimum height=0.55cm, align=center, text width=2.5cm},
  leaf/.style={align=left, font=\scriptsize, anchor=north, text width=2.5cm, inner xsep=0pt},
  every path/.style={draw}
]
\node[dim, minimum width=4.2cm, text width=4.1cm, fill=black!10] (root) at (0,0) {\textbf{Hybrid PQC--QKD design}};
\foreach \x/\n/\t in {-10.15/d1/D1 Integration layer, -7.25/d2/D2 Combiner, -4.35/d3/D3 Authentication scope \& source, -1.45/d4/D4 QKD topology, 1.45/d5/D5 Trusted entities, 4.35/d6/D6 Key-delivery interface, 7.25/d7/D7 Classical part, 10.15/d8/D8 Evidence (multi-valued)}
  \node[dim] (\n) at (\x,-1.4) {\t};
\foreach \n in {d1,d2,d3,d4,d5,d6,d7,d8} \draw (root.south) -- ++(0,-0.5) -| (\n.north);
\node[leaf] at ([yshift=-0.15cm]d1.south) {AKE (stand-alone)\\TLS 1.3\\IKEv2 / IPsec\\MACsec\\Optical layer / KMS};
\node[leaf] at ([yshift=-0.15cm]d2.south) {XOR\\XOR + IT one-time MACs\\KDF over concatenation\\Chained KDF\\Dual-PRF / ciphertext-bound\\Protocol PSK mixing};
\node[leaf] at ([yshift=-0.15cm]d3.south) {Scopes: endpoint exchange (E), QKD channel (Q), KMS/relay channels (K)\\Sources: PSK, PQ signature, classical signature, KEM-based, assumed, n.r.};
\node[leaf] at ([yshift=-0.15cm]d4.south) {Point-to-point span\\Trusted relay chain\\Untrusted relay (MDI/TF)\\Multi-path / networked\\Abstract (proofs)};
\node[leaf] at ([yshift=-0.15cm]d5.south) {None\\Trusted relay nodes\\KMS\\SDN controller\\In-network combiner};
\node[leaf] at ([yshift=-0.15cm]d6.south) {ETSI 014 (stateless)\\ETSI 004 (stateful)\\ETSI 020 (inter-KMS)\\Proprietary\\Abstract model (proofs)};
\node[leaf] at ([yshift=-0.15cm]d7.south) {Two-way (PQC + QKD)\\Three-way (ECDH + PQC + QKD)};
\node[leaf] at ([yshift=-0.15cm]d8.south) {Game-based proof\\Composable proof\\Formal-verification model\\Prototype measurement\\Testbed / field deployment\\Standard\\Informal analysis};
\end{tikzpicture}}
\caption{Taxonomy of hybrid PQC--QKD designs. D1, D2, D4--D7 take one value per design; D3 takes a value per authentication scope, and D8 lists every evidence type that applies. Tables~\ref{tab:taxonomy_matrix} and~\ref{tab:taxonomy_matrix_b} place the surveyed designs.}
\Description{A tree with a root node labelled Hybrid PQC-QKD design and eight child dimensions in one row: integration layer, combiner construction, authentication scope and source, QKD topology, trusted entities, key-delivery interface, classical component, and evidence, each listing its categories beneath it.}
\label{fig:taxonomy}
\end{figure*}
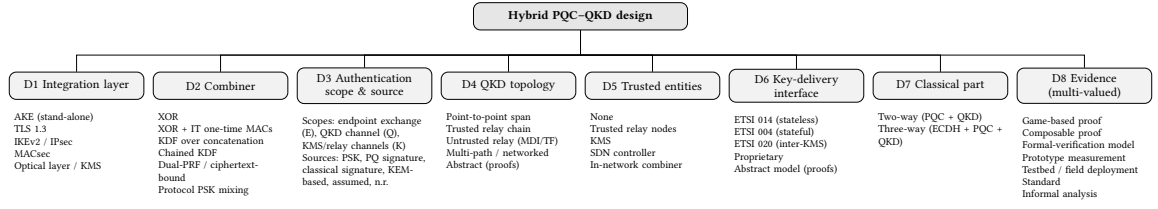

Published designs differ along eight dimensions (Figure~\ref{fig:taxonomy}). We define each and then place the surveyed designs in the matrix of Tables~\ref{tab:taxonomy_matrix} and~\ref{tab:taxonomy_matrix_b}.

\begin{description}[leftmargin=*,labelwidth=!]
  \item[D1 Integration layer.] Where the combined key is consumed: an authenticated key exchange protocol studied on its own (AKE), TLS~1.3, IKEv2/IPsec, MACsec (layer~2), the optical transport layer (OTN/DWDM encryptors), or a key-management system (KMS) that serves keys to any of these.
  \item[D2 Combiner construction.] How the key sources are merged: XOR; XOR with information-theoretic one-time-MAC binding of the transcript and key identifier (as in H\"ovelmanns et al.~\cite{hoevelmanns2025qkdoracles}); a KDF over the concatenation of shared secrets (as in NIST SP~800-56C~Rev.\,2 and ETSI TS~103~744); a chained or nested KDF in which each stage's output keys the next (as in Muckle); a dual-PRF or split-key combiner with ciphertext binding (as in the KEM-combiner literature); or a pre-shared-key mixing mechanism defined by the protocol (RFC~8784/9867).
  \item[D3 Authentication scope and source.] Which channel is authenticated, and by what. We distinguish three scopes: (E) the endpoint key exchange whose key is consumed; (Q) the QKD classical (post-processing) channel between QKD modules; and (K) the key-delivery and key-relay channels between endpoints, KMSs, and relay nodes. For each scope the source is a pre-shared symmetric key (possibly QKD-derived), PQ signatures with a PQ PKI, classical signatures, KEM-based authentication, assumed by the model, or not reported (n.r.).
  \item[D4 QKD topology.] How QKD keys reach the endpoints physically: a single point-to-point span, a chain of trusted relay nodes, untrusted-relay protocols (MDI- or TF-QKD) within a span, or multi-path or networked delivery. ``Abstract'' is recorded when a proof does not fix a topology.
  \item[D5 Trusted entities.] Which entities besides the two endpoints must be trusted with QKD key material: none, trusted relay nodes, the KMS at each site, an SDN controller or orchestrator with key access, or an in-network combiner (X.1714).
  \item[D6 Key-delivery interface.] How the endpoint obtains the QKD key: ETSI GS QKD 014 (stateless REST with key identifiers), ETSI GS QKD 004 (stateful), ETSI GS QKD 020 (inter-KMS), a proprietary interface, or, in proofs, an abstract model (shared array, oracle, or explicitly modelled QKD phase).
  \item[D7 Classical component.] Whether a classical ECDH/DH contribution is retained alongside PQC and QKD (three-way hybrid) or omitted (two-way hybrid).
  \item[D8 Evidence.] What supports the design: a game-based proof, a composable proof, a formal-verification (symbolic) model, a prototype implementation with measurements, a testbed or field deployment, a standard, or informal analysis. A study can carry several evidence types (for example, a proof and an implementation), and we list all that apply.
\end{description}

The dimensions are orthogonal: a chained-KDF combiner (D2) can appear at the AKE, TLS, or MACsec layer (D1) with PSK or signature authentication (D3), over any topology (D4). Separating D4--D6 matters because they fail differently: topology determines whether relays exist, trusted entities determine who holds keys in the clear, and the delivery interface determines how a key is bound to a session. Two dimensions deserve further comment. On D2, the distinction that matters for security is not XOR versus KDF but whether the combiner binds the public values (ciphertexts, key identifiers, transcript) so that an adversary who controls one input cannot influence the output in a way that depends on the other; this is the lesson of the KEM-combiner literature~\cite{giacon2018kem,bindel2019hybrid} and of the dependent-key attacks of H\"ovelmanns et al.~\cite{hoevelmanns2025qkdoracles}, and it is developed in Section~\ref{sec:combiners}. On D3, a design that authenticates the endpoint exchange with PQ signatures gives up the unconditional security of the key against an adversary who can forge those signatures \emph{during} the exchange, but retains it against an adversary who breaks them only later; a design that authenticates with pre-shared keys avoids this at the cost of key pre-distribution. Muckle and Muckle+ sit on opposite sides of exactly this choice~\cite{dowling2020muckle,bruckner2023mucklePlus}. Authentication of the endpoint exchange says nothing about the QKD and KMS channels (scopes Q and K), which the proofs assume and only some deployments document.

\begin{table*}[t]
\caption{Surveyed designs placed in the taxonomy (part 1: protocol and combiner designs). Columns follow Figure~\ref{fig:taxonomy}. D3 gives the authentication source for the endpoint exchange (E), the QKD classical channel (Q), and the key-delivery and relay channels (K); ``assumed'' means the proof takes the channel as authenticated. D8 lists every evidence type that applies; proofs are expanded in Table~\ref{tab:proofs}. n.r.\ = not reported; n.s.\ = not specified by the model. Entries marked $^\ast$ restate findings of H\"ovelmanns et al.~\cite{hoevelmanns2025qkdoracles}.}
\label{tab:taxonomy_matrix}
\scriptsize\setlength{\tabcolsep}{3pt}
\resizebox{\textwidth}{!}{%
\begin{tabular}{@{}L{2.3cm}L{1.3cm}L{2.2cm}L{2.4cm}L{1.5cm}L{1.4cm}L{1.5cm}L{1.1cm}L{2.4cm}@{}}
\toprule
\textbf{Design} & \textbf{D1 Layer} & \textbf{D2 Combiner} & \textbf{D3 Auth.\ (E / Q / K)} & \textbf{D4 Topology} & \textbf{D5 Trusted} & \textbf{D6 Delivery} & \textbf{D7 Classical} & \textbf{D8 Evidence} \\
\midrule
Muckle~\cite{dowling2020muckle} (2020) & AKE & Chained KDF (HKDF stages) & E: PSK (MAC); Q, K: assumed & Abstract & n.s. & Abstract (shared array; key fixed per stage$^\ast$) & Three-way & Game-based proof (HAKE) + implementation benchmarks (LAN/WAN) \\
Muckle+~\cite{bruckner2023mucklePlus} (2023) & AKE & Chained KDF & E: PQ signature + MAC; Q, K: assumed & Multi-path over trusted nodes (field: 3 links, 2 disjoint paths) & Trusted nodes, KMS & ETSI 014 (implementation); abstract in proof & Three-way & Game-based proof (HAKE) + field implementation on a small QKD network \\
Garms et al.\ (Muckle++)~\cite{garms2024experimental} (2024) & AKE on QKD hardware & Chained KDF with a ``QPRF'' (HKDF output XORed with quantum key material)$^\ast$ & E: MAC and/or PQ signatures; Q, K: n.r. & Point-to-point & n.r. & Integrated: QKD post-processing on the same FPGA & Three-way & Security analysis (status disputed$^\ast$) + implementation on a commercial QKD prototype \\
VMuckle~\cite{Buruaga2025-VMuckle} (2025) & AKE supplying the MKA root key (MACsec) & Modular chained KDF; PSK, PQC, QKD inputs configurable & E: PSK and/or PQ signature; Q, K: assumed & Abstract (QKD simulated in evaluation) & n.s. & Out-of-band GetKey (abstract) & Configurable & Game-based proof (HAKE) + prototype benchmarks (simulated QKD); MadQCI deployment planned \\
Ferreira AKE~\cite{ake2026_pqc_qkd} (2026) & AKE & Extract/Expand key schedule over KEM secret, QKD key, and internal state & E: PQ signatures + AE keyed from internal state; Q: QKD classical data protected by AE and signatures; K: party--QKD-endpoint link assumed trusted & End-to-end link, no trusted nodes & None (endpoint link trusted) & Explicit BB84 phase inside the protocol & PQ KEM, optionally hybrid with classical & Game-based proof (soundness, authentication, key indistinguishability, identity privacy) \\
Battarbee et al.\ (Muckle\#)~\cite{battarbee2025kemauth} (2025) & AKE & Chained KDF (Muckle-style) & E: KEM-based (static KEMs) + MACs; Q, K: assumed & Abstract (QKD simulated in evaluation) & n.s. & Out-of-band GetKey without identifier (abstract) & Three-way & Game-based proof (HAKE) + prototype benchmarks (simulated QKD) \\
Clermont \& Henrich~\cite{clermont2026best} (2026) & AKE for QKD(N) & Multi-input KDF (random-oracle construction, SHA3-512) with context incl.\ identities and QKD key ID & E: KEM-based + MACs; Q: subsumed in black-box QKD; K: n.s. & Neighbouring QKD nodes (direct link) & KMS at each node & GetKey / GetKeyById with QKD key ID (ETSI 014-style) & Two-way & Game-based proof (CK01; ITS in the ROM if QKD does not fail) + benchmarks (simulated KMS) \\
H\"ovelmanns et al.~\cite{hoevelmanns2025qkdoracles} (2025, preprint) & AKE & Triple-KEM key and QKD key; nested IT one-time MACs over transcript incl.\ key ID; session key = XOR & E: KEM-based (static + ephemeral KEMs); Q: assumed (ideal oracle); K: not modelled (single tenancy) & End-to-end link & None beyond endpoints (ideal QKD oracle) & Oracle modelled on ETSI 014 (key IDs) & Two-way & Game-based proof (CK+ with QKD oracles; IT in QKD case, QROM in PQC case) + dependent-key attack on unbound designs \\
Ricci et al.~\cite{ricci2024hybridkeys} (2024) & Key agreement (FPGA) & 3-key combiner (classical KEX + PQ KEM + QKD) & E: KEM-based (as described in~\cite{clermont2026best}) & Point-to-point & n.r. & n.r. & Three-way & IND-CCA claim (disputed$^\ast$) + FPGA implementation and benchmarks \\
Djordjevic~\cite{djordjevic2020joint} (2020) & QKD post-processing & -- (PQC transports reconciliation data, not a second key) & Q: PQC protects reconciliation data & Point-to-point (incl.\ TF-QKD) & n.r. & n.r. & -- & Analysis \\
Aquina et al.~\cite{aquina2024hybrid} (2024) & Review + key agreement & KEM-combiner-style combination of classical, PQC, and QKD keys & E: EdDSA + Dilithium signatures, Poly1305 tags with part of the QKD key$^\ast$ & Point-to-point & n.r. & n.r. & Three-way & Four-page review with a combination proposal; no proof \\
Zeng et al.~\cite{zeng2024practical} (2024, preprint) & Key agreement in KD network & Hybrid KDF over keys from parallel QKD and PQC distribution channels & n.r. & Multi-path / networked & n.r. & n.r. & Two-way & Hybrid designs + network-level security-analysis method; no game-based proof \\
Raj \& Balachandran~\cite{raj2025hybrid} (2025) & Application framework & KDF over X25519 + ML-KEM-768 + QKD key & E: EdDSA / ML-DSA; Q, K: n.r. & n.r. & KMS (Guardian) & ETSI 014 & Three-way & Prototype (short paper) \\
Chen~\cite{Chen2025-NIST-QKD-Hybrid} (2025, preprint) & Key exchange + signature & KDF of ML-KEM and QKD keys & E: ML-DSA / SLH-DSA & n.r. & n.r. & n.r. & Two-way & Informal analysis (preprint) \\
\bottomrule
\end{tabular}}
\end{table*}

\begin{table*}[t]
\caption{Surveyed designs placed in the taxonomy (part 2: protocol integrations, deployments, and standards). Columns and abbreviations as in Table~\ref{tab:taxonomy_matrix}. Rows marked ``no PQC'' document no PQC contribution and are included to separate them from hybrid PQC--QKD designs.}
\label{tab:taxonomy_matrix_b}
\scriptsize\setlength{\tabcolsep}{3pt}
\resizebox{\textwidth}{!}{%
\begin{tabular}{@{}L{2.3cm}L{1.3cm}L{2.2cm}L{2.4cm}L{1.5cm}L{1.4cm}L{1.5cm}L{1.1cm}L{2.4cm}@{}}
\toprule
\textbf{Design} & \textbf{D1 Layer} & \textbf{D2 Combiner} & \textbf{D3 Auth.\ (E / Q / K)} & \textbf{D4 Topology} & \textbf{D5 Trusted} & \textbf{D6 Delivery} & \textbf{D7 Classical} & \textbf{D8 Evidence} \\
\midrule
Rubio Garc\'ia et al.~\cite{garcia2023camad,jsac2025_hybrid} (2023, 2025) & TLS 1.3; IPsec; MACsec & KDF over concatenated secrets & E: TLS certificates (classical/PQ); K: n.r. & Experimental network testbed & KMS & ETSI 014 & Three-way & Prototype measurements \\
QKD-KEM / OpenSSL provider~\cite{blanco2025qkdkem} (2025) & TLS 1.3 & PQ shared secret $\|$ QKD key inside one KEM operation; TLS key schedule & E: TLS certificates; K: HTTPS between KMS and endpoints (proof of concept) & Point-to-point (Vigo dark fibre) & KMS (remote from endpoints) & ETSI 004 and 014 & Two-way & Prototype on ID Quantique Cerberis XGR hardware \\
Hybrid IPsec~\cite{hybrid_ipsec2025} (2025, preprint) & IKEv2/IPsec (strongSwan) & Identifier-based pure-QKD mode; QKD-KEM composed with ML-KEM (parallel) vs.\ sequential RFC 9370 & E: IKEv2 auth; K: n.r. & QKD hardware link (ID Quantique) with a Docker-based test framework & KMS & ETSI 004 and 014 & Two-way & Prototype on ID Quantique hardware \\
Alia et al.~\cite{alia2024ipsec} (2024, preprint) --- classical + QKD, no PQC & IPsec (firewalls) & IKEv2 prf+ over SK\_d (from the DH-based IKE SA) and the QKD key in a modified Child SA derivation & E: IKEv2 (method n.r.); K: ETSI 014 between the Clarion KX KMS and the firewalls & Point-to-point, 46 km deployed fibre & KMS & ETSI 014 & Classical DH + QKD (no PQC) & Field deployment (45 days; 24 h stress test) \\
Buruaga et al.~\cite{buruaga2025tls} (2025) & TLS in SDN & KDF & E: TLS certificates; K: n.r. & Networked, SDN-controlled & KMS, SDN controller & Key-provisioning system of the SDN-QKD network & Configurable & Prototype/testbed \\
Kozlovi\v{c}s et al.~\cite{kozlovics2023qkdaas} (2023) --- no PQC documented & TLS & ``Virtual KEM'' backed by a QKD service & E: TLS certificates & n.r. & QKD service provider & QKD-as-a-service API & -- (QKD-backed KEM; no PQC component described in the sources we examined~\cite{kozlovics2023qkdaas,blanco2025qkdkem}) & Prototype \\
Wang et al.~\cite{wang2021npjqi}; Yang et al.~\cite{yang2021pqcauth} (2021) & QKD network (authentication) & -- (PQ signatures authenticate the QKD classical channel) & Q: PQ signature (Aigis-Sig) with PKI & Metro; optical switching instead of trusted relays & n.r. & n.r. & -- & Field network (14 user nodes, 5 switching nodes, 36-day run) \\
Berlin OpenQKD hybrid KMS~\cite{geitz2023hybrid,doering2022pqc5g} (2022--2023) & KMS & PQC KEM-protected key delivery; PQC key exchange as an additional key source in the quantum layer & K: PQ signature trust chain + PSK & Trusted-node metro (star) & Trusted nodes, KMS, HSM & ETSI 014 to encryptors & Configurable & Testbed deployment \\
QuLore~\cite{sanz2026quloreadaptivesecurityframework} (2025, preprint) & KMS / control plane (vKMS + SDN controller) & Adaptive per session: pure QKD, KDF over relayed QKD key and ML-KEM secret, or PQC-only & K: PQC-KEM + OTP relay of QKD keys to nodes without QKD & Networked & vKMS, SDN controller & ETSI 014 / 020 & Configurable & Containerised testbed (Kubernetes, ML-KEM, simulated QKD) \\
RFC 8784 / 9867 PPK~\cite{rfc8784,rfc9867} & IKEv2 & PPK mixed into key derivation & E: IKEv2 auth, with the PPK mixed into the AUTH keys SK\_pi/SK\_pr (RFC 8784) & Any & n.s. & Any PSK source (QKD explicitly contemplated) & Per RFC 9370 & Standard \\
ETSI TS 103 744~\cite{etsi_ts103744} & Generic & Concatenation / cascade KDF & -- & Any & -- & -- & Any & Standard \\
ITU-T X.1714~\cite{itut_x1714} & QKDN key supply & Key combination in QKDN (no context input$^\ast$) & -- & QKDN & KMS / in-network combiner & QKDN-internal & Any & Standard \\
\bottomrule
\end{tabular}}
\end{table*}

Four observations follow from Tables~\ref{tab:taxonomy_matrix} and~\ref{tab:taxonomy_matrix_b}. First, the designs with proofs (the Muckle family, Battarbee et al., Ferreira, Clermont and Henrich, H\"ovelmanns et al.) sit at the AKE layer with an abstract key-delivery model (D6), whereas the designs with testbed evidence sit at the TLS/IPsec/MACsec or KMS layers and obtain keys over ETSI~014, 004, or (in QuLore's case) 020. In our corpus, Muckle+ (field implementation on a small QKD network) and the Muckle++ integration of Garms et al.\ (commercial QKD prototype) carry both proof and implementation evidence on QKD hardware, while VMuckle, Muckle\#, and Clermont and Henrich add prototype benchmarks with simulated QKD; H\"ovelmanns et al.\ and Clermont and Henrich are the proofs that bind the QKD key identifier to the session, the former inside a model that also lets the adversary leak or set QKD keys; Zeng et al.~\cite{zeng2024practical} pair network-level hybrid designs with a security-analysis method for key-distribution networks rather than a game-based proof. Second, several works described as ``hybrid PQC--QKD'' in the literature use PQC only to authenticate the QKD classical channel or to protect the key-management plane~\cite{wang2021npjqi,yang2021pqcauth,geitz2023hybrid}, or to transport QKD post-processing data~\cite{djordjevic2020joint}, and Alia et al.~\cite{alia2024ipsec} and Kozlovi\v{c}s et al.~\cite{kozlovics2023qkdaas} document no PQC component; these are legitimate design points, but they are not combinations of two key sources, and the taxonomy separates them (D2 empty, or D3 scope Q). A related design point, introduced by QuLore~\cite{sanz2026quloreadaptivesecurityframework} and implicit in Zeng et al.~\cite{zeng2024practical}, is \emph{adaptivity}: the network selects per session among pure-QKD, hybrid, and pure-PQC modes according to node capability and path availability, which makes the combiner a policy decision of the key-management plane rather than a property of the protocol. Third, separating the authentication scopes shows that the proofs authenticate the endpoint exchange (scope E) and assume the QKD and key-management channels (scopes Q and K), with the exception of Ferreira, whose protocol also protects the classical data of its embedded QKD phase; these channels are documented only in deployment works such as Jinan, Berlin, and QuLore. Fourth, we found no formal-verification (symbolic) model for a design at the MACsec or optical layer, and the proofs we reviewed address shared-infrastructure threats only partially (Section~\ref{sec:threats}).

\section{Key Combination and Security Arguments}\label{sec:combiners}

\subsection{Combiner constructions}\label{sec:combiner_constructions}

Let $k_1,\dots,k_m$ be shared secrets from $m$ sources (classical ECDH, one or more PQ KEMs, one or more QKD keys). The combiner produces the session key $K$. Table~\ref{tab:combiners} compares the constructions found in the surveyed designs and standards.

\paragraph{XOR.} If one input $k_i$ is uniform and independent of everything else the adversary sees, then $K=\bigoplus_j k_j$ is uniform and independent of the remaining inputs: this is the one-time-pad argument, and it is why XOR ``preserves'' the information-theoretic secrecy of a QKD key when the other inputs are merely computational. The argument is correct but narrow. It requires equal-length inputs; it provides no binding to the public values that produced the inputs; and, as a \emph{KEM} combiner, plain XOR of decapsulated keys does not preserve IND-CCA security in general, because an active adversary who controls one ciphertext can steer the corresponding key~\cite{giacon2018kem}. In a hybrid PQC--QKD setting the analogous concern is a compromised or adversarial QKD key-delivery interface that returns an attacker-chosen ``QKD key'': XOR then leaks nothing about the PQC key, but the session key is now a function of an attacker-known value and a value the attacker may be able to influence through the KEM ciphertext. Whether this matters depends on the surrounding protocol; the point is that XOR's information-theoretic property says nothing about it. H\"ovelmanns et al.~\cite{hoevelmanns2025qkdoracles} make the hybrid version of this concrete. In a protocol that outputs $k_{\mathrm{pqc}}\oplus k_{\mathrm{qkd}}$ without binding the QKD key identifier to the session, an active adversary who swaps the key identifiers of two sessions makes those sessions non-matching, may then reveal their keys legitimately, and can combine the revealed keys to recover the target session key (a \emph{dependent-key attack}). Their protocol prevents this by authenticating the transcript, including the key identifier, with nested one-time MACs keyed by parts of the QKD and PQC keys, and outputs the XOR of the remaining key parts; this preserves information-theoretic key secrecy when QKD is secure (Theorem~5.3 in~\cite{hoevelmanns2025qkdoracles}). XOR is therefore not insecure in itself: what matters is whether the inputs and identifiers are bound to the session, and, when information-theoretic key secrecy is the goal, whether that binding uses only information-theoretic primitives.

\paragraph{KDF over concatenation.} NIST SP~800-56C~Rev.\,2~\cite{nist2020sp80056c} permits a hybrid shared secret $Z\|T$ where $Z$ is from an approved scheme and $T$ is an auxiliary secret; ETSI TS~103~744~\cite{etsi_ts103744} defines concatenation and cascade constructions; and the TLS hybrid design~\cite{rfc9954} concatenates shared secrets and relies on the TLS key schedule (HKDF, with transcript binding) for combination. Security follows if the KDF is a dual PRF, that is, a PRF when keyed by \emph{either} input, which HKDF-Extract satisfies under standard assumptions on HMAC; Bindel et al.~\cite{bindel2019hybrid} give the general hybrid-KEM and hybrid-AKE framework, and Aviram et al.~\cite{aviram2022combiners} give combiners with one-wayness-based guarantees and TLS applications. When the QKD key is one of the concatenated inputs, the combined key inherits computational security from the PQC input if QKD fails, and remains computationally indistinguishable from random (though no longer information-theoretically so) if PQC fails; the KDF itself is the assumption that remains. This loss matters only for properties that depend on the key being information-theoretically secret (Section~\ref{sec:secprops}). Application-data confidentiality does not depend on it in the surveyed integrations, because the data are protected by a computational AEAD in either case. Key secrecy does: with a KDF it rests on the KDF's PRF assumption even when QKD is secure. H\"ovelmanns et al.\ argue that if the combiner is only computationally secure, a system built from computational primitives alone may offer the same guarantee at lower complexity~\cite{hoevelmanns2025qkdoracles}; this is part of the value question in Section~\ref{sec:open}.

\paragraph{Chained KDF (Muckle).} Muckle~\cite{dowling2020muckle} derives a chain of keys in which the output of one HKDF stage keys the next, with classical, PQ, and QKD secrets entering at successive stages. Its main theorem bounds key indistinguishability when the PQ KEM or the QKD input is uncompromised; because the derivation uses a PRF, the guarantee is computational even when QKD is secure~\cite{dowling2020muckle,hoevelmanns2025qkdoracles}. H\"ovelmanns et al.\ also note that the HAKE model fixes the QKD key per session stage, so it does not capture an adversary who causes a party to use the wrong QKD key~\cite{hoevelmanns2025qkdoracles}. Muckle+~\cite{bruckner2023mucklePlus} keeps the chain and replaces PSK-based authentication with PQ signatures, which required non-trivial changes to the proof; H\"ovelmanns et al.\ report an issue in how the Muckle+ analysis defines the tested session key, which they state was resolved in VMuckle. VMuckle~\cite{Buruaga2025-VMuckle} modularises the chain so that inputs can be configured per deployment (PSK-only, PQC+PSK, PQC+QKD) and uses its output as the master session key from which MKA derives the MACsec key hierarchy, leaving IEEE 802.1X and 802.1AE unchanged.

\paragraph{Protocol-defined PSK mixing.} RFC~8784~\cite{rfc8784} mixes a post-quantum pre-shared key (PPK) into IKEv2's key derivation; the RFC explicitly names QKD as a PPK source. RFC~9867~\cite{rfc9867} moves PPK mixing into the IKE\_INTERMEDIATE and CREATE\_CHILD\_SA exchanges so that the initial IKE SA is also protected and so that frequently refreshed PPKs, again explicitly QKD keys, can be used without tearing down the SA. RFC~8784 mixes the PPK not only into SK\_d but also into SK\_pi and SK\_pr, the keys used to compute the IKE\_AUTH payloads, so that both sides detect a PPK mismatch cleanly; an initiator without the right PPK therefore fails authentication. The PPK does not replace the IKEv2 authentication method, but authentication is not independent of it. RFC~8784 also defines an optional PPK-only authentication mode and a fallback (NO\_PPK\_AUTH) whose authentication data are computed without the PPK~\cite{rfc8784}. RFC~9867 instead computes SKEYSEED$' = \mathrm{prf+}(\mathrm{PPK}, \mathrm{SK\_d})$ after the IKE\_INTERMEDIATE exchange and recomputes all SK\_* keys from it, including SK\_pi and SK\_pr, which are then used in IKE\_AUTH (its Section~3.1.1); its security considerations note that the PPK is therefore used in authentication too, so the exchange is quantum-secure even against an active attacker (Section~4). For CREATE\_CHILD\_SA it mixes a fresh PPK into SK\_d (Section~3.2.1)~\cite{rfc9867}. RFC~9370~\cite{rfc9370} adds ML-KEM (or any additional KEM) as further key exchanges. Together these define the standardised way to build a three-way hybrid in IPsec without any new combiner: the IKEv2 PRF chain does the combination. ITU-T X.1714~\cite{itut_x1714} defines key combination inside a QKD network's key-supply layer, so that the key handed to an application may already be a combination of a QKD key and other material; H\"ovelmanns et al.\ note that this combiner takes no ciphertexts or other context as input and that X.1714 gives no formal security definition for it~\cite{hoevelmanns2025qkdoracles}.

\begin{table*}[t]
\caption{Key-combiner constructions used in surveyed designs and standards. ``Notion if PQC breaks'' and ``if QKD breaks'' give the security that the combined key retains when the named component fails while the other holds; ``Binding'' states whether public values (ciphertexts, key IDs, transcript) are bound into the derivation.}
\label{tab:combiners}
\footnotesize\setlength{\tabcolsep}{4pt}
\begin{tabular}{@{}L{2.1cm}L{2.6cm}L{2.1cm}L{2.1cm}L{1.4cm}L{3.2cm}@{}}
\toprule
\textbf{Construction} & \textbf{Where used} & \textbf{Notion if PQC breaks} & \textbf{Notion if QKD breaks} & \textbf{Binding} & \textbf{Caveats} \\
\midrule
XOR of keys & Aquina et al.~\cite{aquina2024hybrid}; VMuckle XOR mode~\cite{Buruaga2025-VMuckle}; compared in Ricci et al.~\cite{ricci2024hybridkeys} & Information-theoretic (one-time-pad argument) & Computational, \emph{only if} the QKD input cannot be attacker-influenced & None & Not IND-CCA-preserving as a KEM combiner~\cite{giacon2018kem}; vulnerable to dependent-key attacks when key IDs are not bound~\cite{hoevelmanns2025qkdoracles}; equal lengths required; no domain separation \\
XOR with IT one-time-MAC binding & H\"ovelmanns et al.~\cite{hoevelmanns2025qkdoracles} & Information-theoretic (Thm.~5.3 in~\cite{hoevelmanns2025qkdoracles}) & Computational (QROM; weak forward secrecy) & Yes (transcript incl.\ key ID, ciphertexts, party IDs, via nested one-time MACs) & Honest QKD keys modelled as perfectly uniform; end-to-end QKD; preprint \\
Multi-input KDF with context (random-oracle construction) & Clermont \& Henrich~\cite{clermont2026best} & Information-theoretic in the ROM if QKD does not fail (Thm.~2 in~\cite{clermont2026best}) & Computational (CK01; Thm.~1) & Yes (sub-protocol transcripts, identities, QKD key ID) & QKD treated as a black box with failure probability $\Pr[\text{QKD fail}]$; ITS relies on the random-oracle model \\
KDF over concatenation (HKDF, KMAC) & TLS hybrid~\cite{rfc9954}; SP 800-56C~\cite{nist2020sp80056c}; ETSI TS 103 744~\cite{etsi_ts103744}; Rubio Garc\'ia et al.~\cite{jsac2025_hybrid}; Chen~\cite{Chen2025-NIST-QKD-Hybrid} & Computational (KDF as PRF keyed by QKD input) & Computational (KDF as PRF keyed by PQC input) & Yes when transcript/context is included & Requires dual-PRF property~\cite{bindel2019hybrid}; ordering and length encoding matter (ETSI TS 103 744 fixes an encoding) \\
Chained KDF & Muckle~\cite{dowling2020muckle}, Muckle+~\cite{bruckner2023mucklePlus}, VMuckle KDF mode~\cite{Buruaga2025-VMuckle} & Computational & Computational & Yes (per-stage context) & Proven in HAKE model; computational at every stage because the derivation uses a PRF; HAKE fixes the QKD key per stage, so key-ID confusion is outside the model~\cite{hoevelmanns2025qkdoracles} \\
Dual-PRF / one-wayness-based combiners & Bindel et al.~\cite{bindel2019hybrid}; Aviram et al.~\cite{aviram2022combiners} & Computational & Computational & Yes (ciphertexts) & Designed for KEM$\times$KEM; QKD is not a KEM, and H\"ovelmanns et al.\ argue these combiners do not accept a QKD input directly~\cite{hoevelmanns2025qkdoracles} \\
PPK mixing in IKEv2 & RFC 8784, RFC 9867~\cite{rfc8784,rfc9867} with RFC 9370~\cite{rfc9370} & Computational (IKEv2 PRF keyed by PPK) & Computational (IKEv2 PRF keyed by KE outputs) & Yes (IKEv2 nonces and SPIs) & RFC 8784 does not protect the initial IKE SA; RFC 9867 does; in RFC 8784 the PPK also enters SK\_pi/SK\_pr and thus authentication \\
QKDN-internal combination & ITU-T X.1714~\cite{itut_x1714} & Depends on KMS construction & Depends on KMS construction & None (no context input~\cite{hoevelmanns2025qkdoracles}) & Combination happens before the application sees the key; application cannot verify it; no formal security definition \\
\bottomrule
\end{tabular}
\end{table*}

\paragraph{Assessment.} The evidence above suggests that XOR versus KDF is not, by itself, the axis that determines security. The standardised constructions we reviewed (SP~800-56C, ETSI TS~103~744, the TLS hybrid design, and IKEv2 PPK mixing) derive keys with a KDF or PRF over the inputs and the protocol context; X.1714 defines combination inside the QKD network without a context input~\cite{hoevelmanns2025qkdoracles}. A KDF-based combiner gives up information-theoretic key secrecy. This does not change application-data confidentiality in designs that use a computational AEAD, but it does change what key secrecy rests on. An XOR-based combiner preserves information-theoretic key secrecy only if the inputs and the QKD key identifier are bound to the session with information-theoretic primitives, as in H\"ovelmanns et al.~\cite{hoevelmanns2025qkdoracles}. The properties that distinguish combiners are therefore (i) binding of public values, including QKD key identifiers; (ii) behaviour when one input is attacker-influenced rather than merely broken; (iii) whether information-theoretic key secrecy is a goal and is preserved; and (iv) whether the construction has a proof in a model that includes the QKD key-delivery interface. On (iv), the KEM-combiner literature does not model a QKD source. Among the protocol proofs we reviewed, the Muckle family and Battarbee et al.\ model QKD as an abstract shared key source retrieved without an identifier; Ferreira models the BB84 phase itself, including its finite-key security loss; Clermont and Henrich treat QKD as a black box with an ETSI 014-style identifier interface and bind the identifier through the KDF context; and H\"ovelmanns et al.\ model an ETSI 014-like interface with key identifiers and adversarial leakage or overriding of QKD keys (Table~\ref{tab:proofs}). Clermont and Henrich state that the XOR-based combiner of H\"ovelmanns et al.\ introduces a dependent-key vulnerability that a context-binding combiner avoids~\cite{clermont2026best}, whereas H\"ovelmanns et al.\ bind the key identifier with one-time MACs before the XOR~\cite{hoevelmanns2025qkdoracles}; the two claims have not been reconciled in the literature we reviewed.

\subsection{Security arguments}\label{sec:proofs}

Table~\ref{tab:proofs} provides a standardized summary of security proofs for hybrid PQC-QKD systems. Table~\ref{tab:proofs} shows details about each citation regarding security model, assumptions about authentication and key distribution, limitations on attackers' knowledge or abilities in the security model, security properties claimed, other limitations of the proofs, and locations of security proofs within each citation. Proofs of parts of a deployed system, including abstract-model key-exchange proofs, do not provide security proofs for deployed systems; for other security limitations of deployed systems, see Section~\ref{sec:threats}.

The first security analysis of a QKD-AKE protocol by Mosca, Stebila, and Ustao\u{g}lu~\cite{mosca2013qkdake} added an oracle to the eCK model to account for quantum communication and considered potential applications of QKD and classical AKE keys in end-to-end protocols~\cite{hoevelmanns2025qkdoracles}. Other studies on this subject have explored four key topics.

\emph{How QKD is modeled.} HAKE proof variants (Muckle, Muckle+, VMuckle, Battarbee et al.\ proofs) model QKD keys as a shared set of uniformly random keys. H\"ovelmanns et al.\ observe that the HAKE modeling approach binds parties to QKD keys at each stage of the protocol, making it impossible for parties to provide IDs that identify QKD keys and precluding attacks on parties that lead them to use the wrong QKD keys~\cite{hoevelmanns2025qkdoracles}. Ferreira~\cite{ake2026_pqc_qkd} models QKD's BB84 phase as part of the AKE protocol and applies the finite-key results of Tomamichel and Leverrier to model QKD's security loss. Clermont and Henrich~\cite{clermont2026best} model QKD as a black-box SK-secure key-exchange protocol that fails with probability $\Pr[\text{QKD fail}]$ and where parties can retrieve keys using IDs. H\"ovelmanns et al.\ model QKD as an oracle similar to those specified in ETSI GS QKD 014, where an honest party receives a uniformly random QKD key with a single ID, and its peer can use the ID to retrieve the same key; the adversary can leak (QKD-LEAK) or override (QKD-OVERRIDE) single QKD keys. H\"ovelmanns et al.'s QKD oracle models include IDs for QKD keys and thus model the dependent-key attacks in Section~\ref{sec:combiner_constructions}.

\emph{AKE models used.} The majority of papers prove security in the HAKE model of Dowling et al.\ and related models or CK01-style models~\cite{canetti2001analysis,clermont2026best}. H\"ovelmanns et al.\ integrate their QKD oracle into the CK+ model~\cite{krawczyk2005hmqv} using pseudo-code for CK+ and prove security for a new protocol that combines a triple-KEM handshake with QKD. They also prove security of the PQC protocol in the quantum random oracle model and information-theoretic key secrecy for the QKD protocol (Theorem~5.3 in~\cite{hoevelmanns2025qkdoracles}).

\emph{Relation to composability.} Security definitions for composable QKD protocols appear in the abstract-cryptography and universal-composability literature~\cite{portmann2022security,canetti2001universally,maurer2011constructive}. Game-based and UC-composability proofs are not known for the security of hybrid PQC, QKD protocols. Some composability considerations in game-based QKD+AKE security proofs include Ferreira's inclusion of the QKD protocol's security degradation in proving the security of AKE. With the security parameters of existing or commercially available BB84 implementations, on the order of $10^{-9}$ to $10^{-10}$, Ferreira shows that the game-based security bound for their hybrid protocol is not meaningful over the number of expected protocol runs for an installed protocol, rendering the QKD security bound tighter than computational bounds~\cite{ake2026_pqc_qkd}. H\"ovelmanns et al. examine the binding of QKD keys to AKE sessions through key identifiers but do not model $\varepsilon$-secure QKD keys and partially model perfectly uniform QKD keys. None of the game-based security proofs we reviewed modeled trusted relays or inter-KMS relays. Composing the QKD bound $\varepsilon_{\mathrm{QKD}}$ with the key-management layer and AKE advantage remains an open problem (O2 in Section~\ref{sec:open}).

\emph{Authentication.} Muckle uses pre-shared keys and Muckle+ PQ signatures and authentication schemes based on KEMs developed by Battarbee et al.\ \cite{battarbee2025kemauth} and Clermont and Henrich~\cite{clermont2026best}. If the QKD part is secure, these schemes have information-theoretic security guarantees in the unconditional security model. H\"ovelmanns et al.\ use static KEMs and nested one-time MACs for authentication and to bind the QKD and PQCrypto parts of their system.

H\"ovelmanns et al.\ discuss other issues in previous work. Other issues mentioned by H\"ovelmanns et al.\ are included in Table~\ref{tab:proofs} as issues raised by H\"ovelmanns et al., not us. Other issues include flaws in the session-key definitions used in Muckle+ testing and a fix in VMuckle, flaws in the dual PRF representation of a base key-derivation algorithm used in Muckle++ proofs by Garms et al.~\cite{garms2024experimental} and a claimed lack of information-theoretic security in those proofs, flaws in the 2024 preprint by Battarbee et al., and a flaw in the chosen-ciphertext property claim for the 3-key combiner of Ricci et al.\ and a claim that the chosen-ciphertext property does not apply to an interactive protocol~\cite{hoevelmanns2025qkdoracles}. Two of the other issues identified by H\"ovelmanns et al.\ are addressed in subsequent work. The published Muckle\# preprint provides complete definitions of cleanness and malicious acceptance and clarifies that the concerns raised by H\"ovelmanns et al.\ about Muckle\# apply to an earlier draft; Abelard et al.\ make a similar point~\cite{abelard2026assessment}. The published VMuckle preprint tests application traffic secrets derived from the master secret rather than the master secret itself (Theorem~1)~\cite{Buruaga2025-VMuckle}. Other authors disagree about the status of Muckle++ security proofs. H\"ovelmanns et al.\ and Ferreira cite supporting materials for Muckle++ proofs, while Battarbee et al.\ say there are no Muckle++ security proofs~\cite{hoevelmanns2025qkdoracles,ake2026_pqc_qkd,battarbee2025kemauth}. Ferreira notes that HAKE-based proofs only establish key indistinguishability properties without defining authentication~\cite{ake2026_pqc_qkd}.

\begin{table*}[t]
\caption{Security-proof comparison for hybrid PQC--QKD protocols. For each record, the entries summarise the original study: the security model; assumptions about authentication and key sources; the compromise or leakage the model allows; the guarantee actually established; limitations of the analysis; and the theorem or section supporting the result. A proof of a component does not establish the security of a deployed system (Section~\ref{sec:threats}). Entries marked $^\ast$ restate findings of H\"ovelmanns et al.~\cite{hoevelmanns2025qkdoracles}; ``as cited in'' marks support taken from a secondary source.}
\label{tab:proofs}
\scriptsize\setlength{\tabcolsep}{3pt}
\resizebox{\textwidth}{!}{%
\begin{tabular}{@{}L{1.8cm}L{1.9cm}L{2.6cm}L{2.4cm}L{2.6cm}L{2.8cm}L{1.4cm}@{}}
\toprule
\textbf{Work} & \textbf{Security model} & \textbf{Assumptions (authentication; key sources)} & \textbf{Allowed compromise / leakage} & \textbf{Established guarantee} & \textbf{Limitations} & \textbf{Support} \\
\midrule
Muckle~\cite{dowling2020muckle} & HAKE (game-based, cleanness predicates) & PSK-based MAC authentication; ECDH + PQ KEM + QKD key, QKD as a shared array with a fixed key per stage$^\ast$ & HAKE queries: session-key reveal, corruption of long-term keys, compromise of ephemeral classical, PQ, and QKD keys and of the session state, subject to the cleanness predicate & Computational key indistinguishability if the PQ KEM or QKD input is uncompromised; forward and post-compromise security & Authentication lost if PSK compromised; wrong-key (key-ID) attacks outside the model$^\ast$; desynchronisation by dropping one message$^\ast$; case bounds added rather than minimised$^\ast$ & Thm.~1 in~\cite{dowling2020muckle} \\
Muckle+~\cite{bruckner2023mucklePlus} & HAKE with predicate clean$_{\text{Muckle+}}$ & PQ signatures + MAC; dual PRF, EUF-CMA MAC and signature, IND-CPA KEMs; QKD keys via multi-path distribution (required by the authors, not captured by HAKE$^\ast$) & HAKE queries (no PSK); long-term signature-key corruption & Computational key indistinguishability with signature authentication; forward and post-compromise security & Tested key is the master secret, which makes the stated notion trivially distinguishable$^\ast$~\cite{ake2026_pqc_qkd}; key-delivery interface not modelled & Thm.~1, Sec.~3.3 in~\cite{bruckner2023mucklePlus} \\
Garms et al.\ (Muckle++)~\cite{garms2024experimental} & HAKE & MAC and/or signatures; ``QPRF'' modelled as a dual PRF$^\ast$ & HAKE queries & Computational key indistinguishability; the claimed ITS is not established by a dual-PRF model$^\ast$ & Source of the quantum key material in the QPRF not specified$^\ast$; proof status disputed (Battarbee et al.\ describe no proof~\cite{battarbee2025kemauth}) & Thm.~3, Supporting Information (as cited in~\cite{hoevelmanns2025qkdoracles}); Def.~13 (as cited in~\cite{ake2026_pqc_qkd}) \\
VMuckle~\cite{Buruaga2025-VMuckle} & HAKE with predicates clean$_{\text{VM}}$ and clean$_{\text{cVM}}$ & PSK and/or PQ signature; post-quantum dual PRF, EUF-CMA MAC and signature, IND-CPA KEM; QKD key obtained out of band and treated as uniform & HAKE queries; at least one authentication secret of each party uncorrupted until acceptance & Post-quantum HAKE security of the application traffic secrets; forward and post-compromise security & Hand-off to MKA outside the model; QKD key fixed per stage (HAKE) & Thm.~1, App.~C in~\cite{Buruaga2025-VMuckle} \\
Ferreira~\cite{ake2026_pqc_qkd} & Model with separate soundness, explicit-authentication, key-indistinguishability, and identity-privacy experiments (standard model) & PQ signatures; KEM (IND-CPA/IND-1CCA); AE; BB84 phase with finite-key analysis; party--QKD-endpoint link trusted & Reveal of session keys, long-term keys, ephemeral KEM keys, ephemeral QKD keys, and internal state, per freshness predicates; KCI, forward and post-compromise security & Each property holds if the KEM and signature, the QKD phase, or the internal state is uncompromised, with the QKD security loss in the bound & QKD loss with deployed parameters yields non-meaningful bounds; end-to-end QKD only (no trusted nodes) & Thm.~2, Sec.~5.1; QKD bound Thm.~1, Sec.~4.7 in~\cite{ake2026_pqc_qkd} \\
Battarbee et al.\ (Muckle\#)~\cite{battarbee2025kemauth} & HAKE with predicate clean$_{\text{Muckle\#}}$ & KEM-based authentication (IND-CCA static KEM); dual PRF, EUF-CMA MAC, IND-CPA ephemeral KEMs; QKD key retrieved without identifier & HAKE queries; passive and active cases of the cleanness predicate & Key indistinguishability with implicit KEM authentication made explicit by MACs & Definitions questioned for the 2024 preprint$^\ast$ are stated in full in the published version; QKD interface not modelled & Thm.~1, Sec.~3.1 in~\cite{battarbee2025kemauth} \\
Clermont \& Henrich~\cite{clermont2026best} & CK01 (SK security with forward secrecy); multi-input KDF security of Backendal et al. & KEM-based AKE of Boyd et al.\ (IND-CCA/IND-CPA KEM, EUF-CMA MAC); QKD as a black box with failure probability; SHA3 random-oracle KDF & Corrupt, session-state reveal, session-key reveal, expire-session & SK security bounded by the minimum of the KEM-AKE and QKD terms; ITS in the ROM if QKD does not fail & QKD internals and classical-channel authentication subsumed in the black box; ROM for the ITS result & Thms.~1--2, Sec.~4; Lemma~1, Sec.~2.6 in~\cite{clermont2026best} \\
H\"ovelmanns et al.~\cite{hoevelmanns2025qkdoracles} (preprint) & CK+ with QKD oracles modelled on ETSI 014 & Static KEM key pairs; end-to-end QKD, single tenancy; honest QKD keys perfectly uniform; QROM for the PQC case & Session-key reveal (including non-matching sessions), long-term key corruption, state reveal, QKD-LEAK and QKD-OVERRIDE of individual QKD keys & PQC case: computational key indistinguishability (QROM), weak forward secrecy. QKD case: information-theoretic key indistinguishability. Captures dependent-key attacks & $\varepsilon$-secure QKD and denial of service not modelled; no relays or multi-tenancy; a known PQC state-reveal impersonation attack is excluded from the PQC game & Sec.~4 (model), Sec.~5 and Thm.~5.3 (ITS) in~\cite{hoevelmanns2025qkdoracles} \\
Ricci et al.~\cite{ricci2024hybridkeys} & Combiner analysis (quantum standard model) & Classical KEX + PQ KEM + QKD; uniform QKD key & n/a (combiner, not protocol) & IND-CCA of the 3-key combiner claimed; disputed, and IND-CCA argued not to be the relevant notion$^\ast$ & Focus on benchmarking rather than formal combiner analysis~\cite{clermont2026best}; inputs are not all KEMs$^\ast$ & As summarised in~\cite{hoevelmanns2025qkdoracles}, Sec.~1.2 \\
Mosca et al.~\cite{mosca2013qkdake} & eCK extended with a quantum-communication oracle$^\ast$ & BB84 with a signature-authenticated classical channel; no combination with classical or PQ key exchange & As in eCK, with quantum-channel access & Session-key indistinguishability; security of BB84 from an asymptotic proof~\cite{ake2026_pqc_qkd} & Only key indistinguishability defined; hybrid combination not treated & As summarised in~\cite{ake2026_pqc_qkd}, Sec.~1.2 and~\cite{hoevelmanns2025qkdoracles}, Sec.~1.2 \\
Aquina et al.~\cite{aquina2024hybrid} & Informal (4-page paper) & Classical + PQ KEMs + QKD, KEM-combiner-style; signatures and Poly1305 tags$^\ast$ & -- & No formal guarantee & Not a proof & -- \\
Chen~\cite{Chen2025-NIST-QKD-Hybrid} & Informal analysis (preprint) & ML-KEM + QKD; ML-DSA/SLH-DSA & -- & Sketch only & Not a proof & -- \\
TLS hybrid~\cite{rfc9954} (classical+PQC) & Informal + prior TLS proofs; dual-PRF rationale & ECDH + PQ KEM; certificate authentication & n/a & Key indistinguishability if either input holds (by reference to prior analyses) & QKD not in scope; provides the template QKD designs reuse & Security Considerations of~\cite{rfc9954} \\
RFC 8784 / 9867~\cite{rfc8784,rfc9867} & Security considerations only & IKEv2 authentication; PPK mixed into SK\_d, SK\_pi, SK\_pr (RFC 8784) or into SKEYSEED$'$ before IKE\_AUTH (RFC 9867) & DH/KEM broken by a quantum adversary & Confidentiality of the derived keys if the PPK is secret (argued, not proven); a peer without the PPK fails IKE\_AUTH & No formal proof; RFC 8784 does not protect the initial IKE SA; NO\_PPK\_AUTH fallback computed without PPK & RFC 8784 Sec.~3; RFC 9867 Secs.~3.1.1, 3.2.1, 4 \\
\bottomrule
\end{tabular}}
\end{table*}

\subsection{Shared-infrastructure threats}\label{sec:threats}

The hybrid argument concludes ``secure if either component holds.'' It is silent about adversaries who make \emph{both} component models false at once, and in deployed systems the two components share infrastructure in ways that make this plausible. Table~\ref{tab:threats} lists the mechanisms we found discussed in the deployment and attack literature, the component each affects, whether existing proofs' models cover it, and the mitigations reported. The proofs we reviewed cover these rows at most partially: H\"ovelmanns et al.\ give the adversary leakage and overwriting of individual QKD keys and model key identifiers, Ferreira lets the adversary reveal the ephemeral QKD key, and the HAKE-based proofs let it compromise the QKD key of a stage; these abstract the effect of several rows without modelling their mechanism. This is not a flaw in the proofs, whose models are explicit, but it is the reason a hybrid deployment cannot claim ``secure if either holds'' about its \emph{system} on the strength of a protocol proof alone.

\begin{table*}[t]
\caption{Shared-infrastructure threats to hybrid PQC--QKD deployments. ``In proof model?'' refers to the arguments in Table~\ref{tab:proofs}.}
\label{tab:threats}
\footnotesize\setlength{\tabcolsep}{4pt}
\begin{tabular}{@{}L{2.5cm}L{3.2cm}L{1.5cm}L{2.4cm}L{4.3cm}@{}}
\toprule
\textbf{Mechanism} & \textbf{Effect} & \textbf{Affects} & \textbf{In proof model?} & \textbf{Mitigations reported / applicable standard} \\
\midrule
Key-identifier confusion or replay at the key-delivery interface & A party uses the QKD key of another session; combined with session-key reveals, a dependent-key attack recovers a target key~\cite{hoevelmanns2025qkdoracles} & QKD input and session key & Yes in~\cite{hoevelmanns2025qkdoracles}; key ID bound through the KDF context in~\cite{clermont2026best}; not in HAKE-based proofs & Bind the key ID into the authenticated transcript with an IT one-time MAC~\cite{hoevelmanns2025qkdoracles}; ETSI 014 keys are delivered once per party \\
Trusted relay node compromise & Relayed QKD keys held in the clear; if the node also terminates the PQC exchange or hosts the KMS, both key sources are exposed at one point & QKD and PQC & No (proofs assume end-to-end or abstract QKD) & Physical and procedural node protection~\cite{mehic2020quantum,cao2022evolution}; PQC-protected key relay~\cite{geitz2023hybrid}; MDI/TF-QKD remove trust in the relay's measurement but not in end nodes; ITU-T X.1710~\cite{itut_x1710} \\
Key-management system (KMS) compromise & KMS serving keys over ETSI 014/004 sees every QKD key; if it also derives or stores hybrid keys, the combiner's inputs are exposed & QKD (and PQC if combined in KMS) & Partially: per-key leak/\allowbreak override~\cite{hoevelmanns2025qkdoracles}; the KMS as an entity is not modelled & PQC-authenticated and PQC-encrypted KMS channels~\cite{geitz2023hybrid,yang2021pqcauth}; combination at the endpoint rather than in the KMS; X.1714~\cite{itut_x1714} \\
Unauthenticated or weakly authenticated QKD classical channel & Man-in-the-middle on reconciliation yields attacker-known ``QKD key'' & QKD & Assumed away in most proofs; an attacker-set key is abstracted by QKD-OVERRIDE~\cite{hoevelmanns2025qkdoracles} & PSK + IT-MAC with bootstrapping; PQ signatures for bootstrapping~\cite{alleaume2014using,yang2021pqcauth} \\
QKD device implementation attack (detector blinding, source imperfection) & QKD key known to attacker while QBER looks normal & QKD & Effect abstracted by QKD-LEAK/\allowbreak OVERRIDE~\cite{hoevelmanns2025qkdoracles}; device not modelled & MDI-QKD~\cite{Lo2012-MDI}; countermeasures surveyed in~\cite{xu2020secure}; certification ETSI GS QKD 016~\cite{etsi_qkd_016}, ISO/IEC 23837~\cite{iso23837} \\
Co-located PQC side channel or fault & PQC key recovered from a node that also handles QKD keys & PQC (and QKD by co-location) & Effect abstracted by corruption and state reveal in AKE models; mechanism not modelled & Masked/constant-time implementations~\cite{ravi2024survey,tches2022_masking}; separate hardware for PQC and KMS functions \\
Shared randomness or seed reuse & KEM ephemeral randomness and QKD post-processing randomness drawn from one weak RNG & PQC and QKD & No & Independent certified RNGs; a design requirement in ETSI/ISO QKD evaluation~\cite{iso23837} \\
Supply chain (common vendor of QKD module and PQC accelerator) & Backdoored components in both paths~\cite{acmccs2024_repqc} & PQC and QKD & Effect partly abstracted (corruption, QKD-OVERRIDE); mechanism not modelled & Vendor diversity; certified modules; open implementations \\
Fallback/downgrade & Endpoint silently continues PQC-only when QKD key delivery fails, or vice versa, without policy enforcement & Whichever is dropped & Partially (models fix the configuration) & Explicit negotiation and policy (VMuckle configurations~\cite{Buruaga2025-VMuckle}; RFC 9867 PPK policy~\cite{rfc9867}) \\
\bottomrule
\end{tabular}
\end{table*}

Three consequences follow for design and evaluation. First, the combiner should be evaluated at the endpoint that consumes the key, not inside a KMS, so that a KMS compromise exposes at most the QKD input; X.1714-style in-network combination trades this property for operational convenience. Second, a hybrid deployment's evidence should include, alongside a protocol proof, the device-level certification status of the QKD modules and the implementation-security status of the PQC code, because the proof assumes both. Third, the key-delivery interface should be part of the proof model. H\"ovelmanns et al.~\cite{hoevelmanns2025qkdoracles} show that this is feasible for an ETSI 014-like interface and that it exposes dependent-key attacks on protocols that do not bind key identifiers; in the works we reviewed, the stateful ETSI 004 interface, multi-tenant KMS operation, and inter-KMS relay (ETSI GS QKD 020) have not been modelled.
\section{Protocol-Level Integration}\label{sec:protocols}

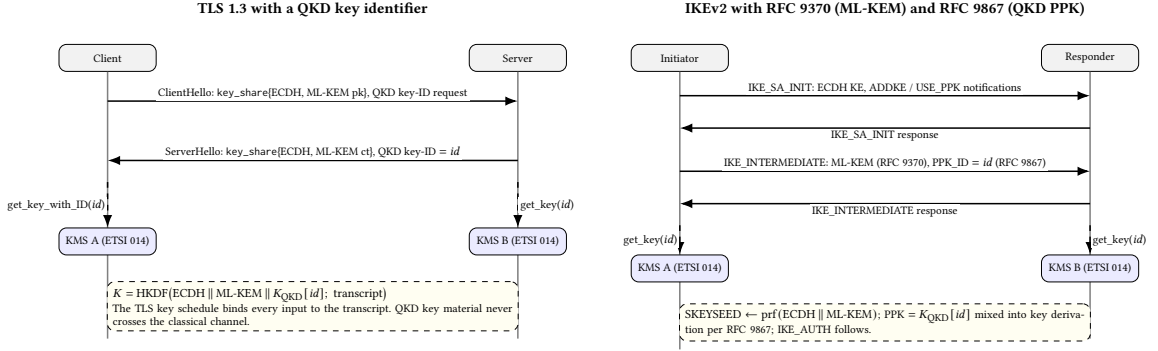
\begin{figure*}[t]
\centering
\resizebox{\textwidth}{!}{%
\begin{tikzpicture}[
  font=\scriptsize,
  >=Latex,
  party/.style={draw, rounded corners, minimum width=1.8cm, minimum height=0.5cm, align=center, fill=black!5},
  kms/.style={draw, rounded corners, minimum width=1.8cm, minimum height=0.5cm, align=center, fill=blue!8},
  msg/.style={->, thick},
  lbl/.style={font=\scriptsize, fill=white, inner sep=1pt},
  note/.style={draw, dashed, rounded corners, align=left, font=\scriptsize, fill=yellow!6, inner sep=3pt}
]
\begin{scope}[xshift=0cm]
\node[font=\small\bfseries] at (3.8,0.9) {TLS 1.3 with a QKD key identifier};
\node[party] (c) at (0,0) {Client};
\node[party] (s) at (7.6,0) {Server};
\draw[thick, gray] (c.south) -- (0,-5.2);
\draw[thick, gray] (s.south) -- (7.6,-5.2);
\draw[msg] (0,-0.8) -- node[lbl, above]{ClientHello: \texttt{key\_share}\{ECDH, ML-KEM pk\}, QKD key-ID request} (7.6,-0.8);
\draw[msg] (7.6,-1.9) -- node[lbl, above]{ServerHello: \texttt{key\_share}\{ECDH, ML-KEM ct\}, QKD key-ID $=\mathit{id}$} (0,-1.9);
\node[kms] (kc) at (0,-3.4) {KMS A (ETSI 014)};
\node[kms] (ks) at (7.6,-3.4) {KMS B (ETSI 014)};
\draw[msg, dashed] (0,-2.3) -- node[lbl, left]{get\_key\_with\_ID($\mathit{id}$)} (kc.north);
\draw[msg, dashed] (7.6,-2.3) -- node[lbl, right]{get\_key($\mathit{id}$)} (ks.north);
\node[note, text width=7.4cm] at (3.8,-4.6) {$K = \mathrm{HKDF}\big(\mathrm{ECDH}\,\|\,\mathrm{ML\text{-}KEM}\,\|\,K_{\mathrm{QKD}}[\mathit{id}];\ \mathrm{transcript}\big)$\\The TLS key schedule binds every input to the transcript. QKD key material never crosses the classical channel.};
\end{scope}
\begin{scope}[xshift=10.6cm]
\node[font=\small\bfseries] at (3.8,0.9) {IKEv2 with RFC 9370 (ML-KEM) and RFC 9867 (QKD PPK)};
\node[party] (i) at (0,0) {Initiator};
\node[party] (r) at (7.6,0) {Responder};
\draw[thick, gray] (i.south) -- (0,-5.4);
\draw[thick, gray] (r.south) -- (7.6,-5.2);
\draw[msg] (0,-0.7) -- node[lbl, above]{IKE\_SA\_INIT: ECDH KE, ADDKE / USE\_PPK notifications} (7.6,-0.7);
\draw[msg] (7.6,-1.3) -- node[lbl, below]{IKE\_SA\_INIT response} (0,-1.3);
\draw[msg] (0,-2.1) -- node[lbl, above]{IKE\_INTERMEDIATE: ML-KEM (RFC 9370), PPK\_ID $=\mathit{id}$ (RFC 9867)} (7.6,-2.1);
\draw[msg] (7.6,-2.7) -- node[lbl, below]{IKE\_INTERMEDIATE response} (0,-2.7);
\node[kms] (ki) at (0,-3.9) {KMS A (ETSI 014)};
\node[kms] (kr) at (7.6,-3.9) {KMS B (ETSI 014)};
\draw[msg, dashed] (0,-3.1) -- node[lbl, left]{get\_key($\mathit{id}$)} (ki.north);
\draw[msg, dashed] (7.6,-3.1) -- node[lbl, right]{get\_key($\mathit{id}$)} (kr.north);
\node[note, text width=7.4cm] at (3.8,-4.9) {$\mathrm{SKEYSEED}\leftarrow\mathrm{prf}(\mathrm{ECDH}\,\|\,\mathrm{ML\text{-}KEM})$; PPK $=K_{\mathrm{QKD}}[\mathit{id}]$ mixed into key derivation per RFC 9867; IKE\_AUTH follows.};
\end{scope}
\end{tikzpicture}}
\caption{Where QKD and PQC enter the handshake. Left: TLS~1.3 designs exchange a QKD key identifier next to the classical and ML-KEM key shares; each side retrieves $K_{\mathrm{QKD}}$ from its local key-management system and the key schedule combines all inputs. Right: IKEv2 uses standardised mechanisms: an additional key exchange (RFC~9370) for ML-KEM and a post-quantum pre-shared key identifier (RFC~9867) for the QKD key. In neither case does QKD key material cross the classical channel.}
\Description{Two message sequence charts. In the TLS 1.3 chart, the client and server exchange ClientHello and ServerHello messages carrying ECDH and ML-KEM key shares and a QKD key identifier, then each retrieves the QKD key from a local key management system before deriving the session key with HKDF. In the IKEv2 chart, the initiator and responder exchange IKE_SA_INIT and IKE_INTERMEDIATE messages carrying ECDH, ML-KEM, and a PPK identifier, then retrieve the QKD key locally and mix it into the IKEv2 key derivation.}
\label{fig:flows}
\end{figure*}

Hybrid PQC--QKD key material reaches applications through existing protocols. Table~\ref{tab:protocol_overview} compares the three layers with published designs; Figure~\ref{fig:flows} shows where the QKD key identifier and the PQC KEM appear in the TLS~1.3 and IKEv2 handshakes. In the integrations we reviewed, the QKD key is obtained from a key-management interface, in most cases ETSI GS QKD 014 (REST, stateless) or 004 (stateful)~\cite{etsi_qkd_014,etsi_qkd_004}; the endpoints exchange a \emph{key identifier} in the handshake and retrieve the corresponding key locally, so that QKD key material does not cross the classical channel. The key identifier is then the only value that binds the QKD key to the session, and H\"ovelmanns et al.\ show that it must be authenticated as part of the transcript~\cite{hoevelmanns2025qkdoracles}.

\begin{table*}[t]
\caption{Hybrid PQC--QKD integration in network protocols. Measurements are given with their metric type (handshake latency, application throughput, QKD secret-key rate, or rekey interval) and the conditions stated by each source; they are not comparable across rows. Table~\ref{tab:measurements} collects them by metric.}
\label{tab:protocol_overview}
\scriptsize\setlength{\tabcolsep}{3.5pt}
\begin{tabular}{@{}L{1.3cm}L{2.9cm}L{2.6cm}L{2.9cm}L{1.6cm}L{2.4cm}@{}}
\toprule
\textbf{Protocol} & \textbf{Mechanism} & \textbf{Governing documents} & \textbf{Reported measurements (metric; conditions)} & \textbf{Fallback} & \textbf{Open issues} \\
\midrule
TLS 1.3 & QKD key ID carried in a \texttt{key\_share} entry (as a named group) or in an extension; ML-KEM (and optionally ECDH) in \texttt{key\_share}; secrets concatenated into the TLS key schedule & RFC 8446~\cite{rfc8446}; RFC 9954, 10024~\cite{rfc9954,rfc10024}; ETSI 014~\cite{etsi_qkd_014}; designs in~\cite{garcia2023camad,jsac2025_hybrid,blanco2025qkdkem,buruaga2025tls} & \emph{Handshake latency}: about $+68\%$ total handshake time for the triple hybrid, about $+104\%$ with an emulated fast ETSI 014 API, in an experimental network scenario~\cite{garcia2023camad}; about $+117\%$ key-establishment cost for hybrid PQC--QKD TLS~\cite{rubio2024comcom}; about 100 ms per ETSI 014 call and full handshakes of 300--350 ms versus 13--18 ms for pure PQC, with QKD nodes in Vigo and endpoints in Madrid~\cite{blanco2025qkdkem}. \emph{Qualitative}: ``non-substantial'' performance decrease and negligible added packet overhead~\cite{jsac2025_hybrid} & Negotiate a PQC-only or classical+PQC group when no QKD key is available & No IANA-registered QKD named group (to our knowledge); key-ID semantics not standardised; no formal analysis of the QKD key-ID exchange \\
IKEv2 / IPsec & PPK (QKD key) mixed into key derivation, including the AUTH keys (RFC 8784), or into IKE\_INTERMEDIATE / CREATE\_CHILD\_SA (RFC 9867); ML-KEM as an additional key exchange (RFC 9370); PPK identifier exchanged, key fetched via ETSI 014 & RFC 8784, 9242, 9370, 9867~\cite{rfc8784,rfc9242,rfc9370,rfc9867}; ETSI 004/014; hybrid prototype~\cite{hybrid_ipsec2025}; measurements in~\cite{jsac2025_hybrid}; classical+QKD deployment (no PQC)~\cite{alia2024ipsec} & \emph{Handshake latency (qualitative)}: sequential RFC 9370 exchanges add one round trip per key exchange; the strongSwan prototype reports that a parallel QKD-KEM composition avoids this and that the identifier-based pure-QKD mode has minimal bandwidth overhead (ID Quantique hardware, ETSI 004/014)~\cite{hybrid_ipsec2025}. Classical+QKD IPsec (no PQC): see Table~\ref{tab:measurements}~\cite{alia2024ipsec} & PPK optional flag (RFC 8784); PQC-only when no PPK & Interoperability profiles binding RFC 9867 PPK IDs to ETSI 014 key IDs not found \\
MACsec (802.1AE) & VMuckle runs as a separate hybrid AKE (PSK and/or PQ signature authentication; ML-KEM and QKD key inputs); its output is the MKA master session key, the root of the MKA key hierarchy; MKA and MACsec run unmodified & IEEE 802.1X/802.1AE~\cite{ieee8021ae}; VMuckle~\cite{Buruaga2025-VMuckle} & \emph{Handshake cost (prototype, simulated QKD)}: about 1.5--2.2~G cycles and 1--84~KB per VMuckle run depending on KEM and signature; 42.2~KB for ML-KEM-1024 + ML-DSA-87 + QKD + ECDH versus 17.29~KB for EAP-TLS, a one-time cost per root key~\cite{Buruaga2025-VMuckle}. Data-plane throughput not measured & PSK-only or PQC+PSK configurations & Hand-off to MKA outside the proof model; no formal-verification model of the VMuckle exchange and its MKA hand-off \\
Optical layer (OTN/DWDM encryptors) & QKD key from KMS feeds the encryptor's key manager; PQC used for KMS authentication or as a second key source & ETSI 014/004; ITU-T Y.3800~\cite{itut_y3800}, X.1714~\cite{itut_x1714}; deployments in~\cite{braun2023berlin,hybrid_brownfield2025} & Vendor-specific & Vendor-specific & Combination inside the KMS is opaque to the endpoint (Section~\ref{sec:threats}) \\
\bottomrule
\end{tabular}
\end{table*}

\subsection{TLS 1.3}

Two integration patterns appear. Rubio Garc\'ia et al.~\cite{garcia2023camad,jsac2025_hybrid} extend the OpenSSL hybrid key exchange so that a QKD key identifier is exchanged alongside the ECDH and ML-KEM key shares; each endpoint retrieves the key over ETSI~014 and the concatenated secrets enter the TLS key schedule, which binds them to the transcript through HKDF. Blanco-Romero et al.~\cite{blanco2025qkdkem} package the same idea as an OpenSSL~3 provider that encapsulates the PQ shared secret and a QKD key identifier in a single KEM operation, with a client-initiated flow that supports the stateful ETSI~004 interface and a server-initiated flow limited to the stateless ETSI~014 interface; Buruaga et al.~\cite{buruaga2025tls} integrate a similar design with SDN-controlled QKD networks and key-provisioning systems, and Kozlovi\v{c}s et al.~\cite{kozlovics2023qkdaas} earlier exposed a QKD service as a ``virtual KEM.'' The two studies that report handshake latency with ETSI-based key retrieval~\cite{garcia2023camad,blanco2025qkdkem} attribute most of the added latency to the QKD key retrieval rather than to ML-KEM (Table~\ref{tab:measurements}); handshake latency and data-plane throughput are different metrics, and the handshake figures cited here say nothing about data-plane throughput. H\"ovelmanns et al.\ note that the CAMAD design carries the QKD key identifier in the ServerHello but gives no security definition or analysis for it~\cite{hoevelmanns2025qkdoracles}; Blanco-Romero et al.\ note that their proof-of-concept carried QKD keys over HTTPS between the key-management entities and the endpoints, which a production deployment must avoid by co-locating endpoints with QKD nodes. Rubio Garc\'ia et al.~\cite{rubio2024comcom} and Buruaga et al.~\cite{buruaga2026benchmark} benchmark hybrid TLS over real infrastructure. To our knowledge, no QKD named group is registered with IANA; the identifiers used in these papers (e.g., \texttt{qkd\_kyber768}) are experimental.

\subsection{IKEv2 / IPsec}

Of the three layers, IPsec is the only one at which standards define a mechanism for each input: RFC~9370 for ML-KEM and RFC~8784/9867 for a QKD-derived PPK. RFC~8784~\cite{rfc8784} mixes a post-quantum pre-shared key into IKEv2 key derivation and names QKD as a source of such keys; its limitations, no protection of the initial IKE SA and no way to use frequently refreshed keys without re-establishing the SA, are addressed by RFC~9867~\cite{rfc9867}, which uses the IKE\_INTERMEDIATE exchange~\cite{rfc9242} and the CREATE\_CHILD\_SA exchange and again names QKD explicitly. RFC~9370~\cite{rfc9370} adds ML-KEM as an additional key exchange in the same IKE\_INTERMEDIATE mechanism. A three-way hybrid (ECDH + ML-KEM + QKD-derived PPK) therefore needs no non-standard combiner: IKEv2's own PRF chain combines the inputs with nonces and SPIs bound in. What is not standardised is the mapping from a PPK identifier to an ETSI~014 key identifier and the policy for PPK refresh. Blanco-Romero et al.~\cite{hybrid_ipsec2025} implement two alternatives in strongSwan, a pure-QKD mode in which an identifier replaces computational key derivation and a QKD-KEM abstraction that composes the QKD key with ML-KEM in parallel, and compare them with the sequential additional-key-exchange pattern of RFC~9370; on ID~Quantique hardware with ETSI~004 and 014 interfaces they find that the parallel composition avoids the per-exchange latency that sequential RFC~9370 exchanges accumulate. Alia et al.~\cite{alia2024ipsec} report QKD-secured IPsec tunnels between two data centres over 46~km of deployed fibre, with keys delivered from the QKD key-management server to the firewalls over ETSI~014. A modified IKEv2 derives each Child SA key with prf+ from SK\_d of the Diffie--Hellman-based IKE SA and the QKD key, so the design combines classical and QKD key material; no PQC component is documented, so we classify this work as QKD-secured IPsec rather than as a hybrid PQC--QKD design (Tables~\ref{tab:taxonomy_matrix_b} and~\ref{tab:testbeds}). Its figures show why metric types must be kept apart: one tunnel reached 80~Gb/s and twelve tunnels an aggregate of 99.62~Gb/s of application throughput, while the QKD link delivered an average secret-key rate of 7.4~kb/s and the keys were refreshed every 120~s~\cite{alia2024ipsec}. The earlier survey by Dervisevic and Mehic~\cite{dervisevic2021ipsec} reviews QKD-in-IPsec proposals that predate the RFCs.

\subsection{MACsec}

VMuckle~\cite{Buruaga2025-VMuckle} is a hybrid AKE that runs before MKA. Its session key, derived from a configurable combination of a pre-shared key, a PQ-signature-authenticated ML-KEM exchange, and a QKD key obtained out of band, is used as the master session key that MKA takes as the authenticated root of its key hierarchy. MKA and the MACsec data plane then run without modification, so VMuckle supplies root key material for MKA without changing IEEE 802.1X or 802.1AE~\cite{Buruaga2025-VMuckle}. Its proof covers the VMuckle exchange (Theorem~1); the hand-off to MKA is outside the model. Its evaluation is a Python prototype with simulated QKD on a notebook, and the authors list deployment on the MadQCI network as future work~\cite{Buruaga2025-VMuckle}. At layer~2 it is thus the design in our corpus with a proof, while Rubio Garc\'ia et al.~\cite{jsac2025_hybrid} report MACsec measurements without a proof. A formal-verification model of the VMuckle exchange together with its hand-off to MKA would be a natural next step.

\subsection{Cross-layer comparison}

Three differences matter. \emph{Where the combiner lives.} TLS and IKEv2 reuse their own key schedules as the combiner, so the binding and dual-PRF properties come from existing analyses; MKA takes an authenticated root key as given and has no key exchange of its own to extend, so VMuckle supplies that root key from a separate AKE. \emph{Where authentication comes from.} TLS designs rely on certificates, which today are classical or classical+PQ; in IKEv2 with RFC~8784, the PPK is mixed into the keys used to compute the AUTH payloads, so successful authentication also requires possession of the PPK, in addition to the configured IKEv2 authentication method (RFC~8784 also discusses PPK-only authentication)~\cite{rfc8784}; VMuckle allows PSK-only authentication of the endpoint exchange, which keeps that exchange free of public-key authentication assumptions. \emph{Where the latency goes.} The TLS studies that report handshake latency attribute most of the added latency to the ETSI~014 round trip~\cite{garcia2023camad,blanco2025qkdkem}, and the IPsec prototype reports the latency effect of sequential versus parallel exchanges~\cite{hybrid_ipsec2025}; in these measurements ML-KEM's contribution is small, consistent with the classical+PQC benchmarks~\cite{paquin2020benchmarking}. Buffering pre-distributed QKD keys in the KMS, which the ETSI interfaces assume and which Clermont and Henrich adopt in their benchmarks~\cite{clermont2026best}, removes QKD key generation from the critical path; the retrieval round trip itself remains.

\section{Deployment Evidence}\label{sec:deployments}

Tables~\ref{tab:testbeds} and~\ref{tab:testbeds_hybrid} list the QKD networks from which the hybrid literature draws operational data, separated by whether PQC was in the key path at all: the networks in Table~\ref{tab:testbeds} are pure-QKD deployments whose data (key rates, availability, trusted-node architecture) constrain hybrid designs but do not test them, and Table~\ref{tab:testbeds_hybrid} lists the testbeds and prototypes that actually combined or authenticated with PQC. In our corpus, the Berlin OpenQKD testbed~\cite{geitz2023hybrid,braun2023berlin,doering2022pqc5g}, the Madrid network (MadQCI) on which the SDN-TLS work was run~\cite{martin2024madqci,buruaga2025tls}, the Eindhoven testbed used for the TLS/IPsec/MACsec measurements~\cite{garcia2023camad,jsac2025_hybrid}, the Vigo--Madrid QKD-KEM and IPsec prototypes~\cite{blanco2025qkdkem,hybrid_ipsec2025}, the Jinan metropolitan network with PQC authentication~\cite{yang2021pqcauth,wang2021npjqi}, and Toshiba's hybrid cryptosystem integration (Muckle++)~\cite{garms2024experimental} report PQC in some part of the key or control path. The deployed-fibre IPsec trial of Alia et al.~\cite{alia2024ipsec} combines classical and QKD keys without a PQC component and is therefore listed with the pure-QKD deployments. Every value in both tables is taken from the cited source; where a source does not state a value the cell reads n.r. Table~\ref{tab:measurements} lists the quantitative results by metric type, with the experimental conditions each source reports.

\begin{table*}[t]
\caption{Pure-QKD networks and deployments whose operational data the hybrid PQC--QKD literature draws on. n.r.\ = not reported in the cited source.}
\label{tab:testbeds}
\scriptsize\setlength{\tabcolsep}{3.5pt}
\begin{tabular}{@{}L{2.1cm}L{1.1cm}L{2.1cm}L{2.0cm}L{2.6cm}L{3.6cm}@{}}
\toprule
\textbf{Network} & \textbf{Years} & \textbf{Scale (as reported)} & \textbf{QKD protocols} & \textbf{PQC in key path?} & \textbf{What the hybrid literature takes from it} \\
\midrule
SECOQC Vienna~\cite{peev2009secoqc} & 2008 & 6 nodes, 8 links, metro & BB84 variants, COW, entanglement-based, CV & No & First multi-vendor trusted-node network; origin of the key-management layering later standardised by ETSI \\
Tokyo QKD Network~\cite{tokyo_qkd_network} & 2010-- & Mesh of 4 access points forming 6 logical nodes; 6 QKD systems; links 1--90 km & Decoy BB84, DPS, COW, entanglement-based (six systems) & No & First GHz-clocked metro network; key relay via trusted nodes; 45 km secure TV conference; three-layer (quantum / key-management / application) architecture \\
Battelle (Ohio)~\cite{morrow2012battelle} & 2012-- & n.r.\ in the cited abstract & Commercial DV-QKD (ID Quantique) & No & Early commercial QKD test bed in the US \\
China backbone and integrated network~\cite{chen2021integrated} & 2017--2021 & Beijing--Shanghai trusted-node backbone (32 nodes, $\sim$2{,}000~km) plus metro networks and satellite links; 4{,}600~km overall & Decoy BB84; satellite & Not in the 2021 report & Scale of trusted-node operation; satellite integration \\
OpenQKD (pan-European)~\cite{OpenQKD2024-D75,openqkd_cordis,brauer2024linking,Ribezzo2023-InterEuropean} & 2019--2023 & Four larger testbeds (Vienna, Madrid, Berlin, Pozna\'n) plus local sites; about 30 use-case demonstrations & DV and CV, multi-vendor & Site-dependent: hybrid PQC--QKD key exchange demonstrated in Berlin & Multi-vendor ETSI 014/004 interoperability; inter-testbed key linking Madrid--Berlin--Pozna\'n~\cite{brauer2024linking}; per-site operational reports in D7.5 \\
Singapore NQSN~\cite{Singapore-NQSN2023} & 2022-- & Star-type QKD network over a Singapore-wide fibre network; layered (quantum / key management / network management) & DV (commercial) & Not in the 2023 report: PQC integration is described as planned & Government and finance use cases; quantum-safe VPN and data-centre trials \\
ORNL CV-QKD field test~\cite{williams2024field} & 2024 & Deployed fibre & CV & No & CV-QKD performance in deployed fibre \\
QKD-secured IPsec trial (Alia et al.)~\cite{alia2024ipsec} & 2024 & Point-to-point, 46 km deployed fibre between two data centres in Singapore; 45-day run & Decoy-state BB84 (ID Quantique Clavis XGR) & No (classical DH + QKD in IKEv2) & QKD-keyed IPsec at up to 80 Gb/s per tunnel and 99.62 Gb/s aggregate over 12 tunnels; secret-key rate 7.4 kb/s; 120 s key refresh; ETSI 014 to firewalls (Table~\ref{tab:measurements}) \\
\bottomrule
\end{tabular}
\end{table*}

\begin{table*}[t]
\caption{Testbeds and prototypes that ran PQC in the key or control path. n.r.\ = not reported in the cited source.}
\label{tab:testbeds_hybrid}
\scriptsize\setlength{\tabcolsep}{3.5pt}
\begin{tabular}{@{}L{2.1cm}L{1.1cm}L{2.1cm}L{2.0cm}L{2.6cm}L{3.6cm}@{}}
\toprule
\textbf{Network} & \textbf{Years} & \textbf{Scale (as reported)} & \textbf{QKD protocols} & \textbf{PQC in key path?} & \textbf{What the hybrid literature takes from it} \\
\midrule
Jinan metro network with PQC authentication~\cite{yang2021pqcauth,wang2021npjqi} & 2021 & 14 user nodes and 5 optical switching nodes; 36-day continuous run & Decoy BB84 (commercial) & Yes: lattice-based PQ signatures (Aigis-Sig) with PKI authenticate the classical channel & PQC as the authentication bootstrap for QKD; optical switching in place of trusted relays (D3 in Table~\ref{tab:taxonomy_matrix_b}) \\
Berlin OpenQKD (Deutsche Telekom)~\cite{geitz2023hybrid,braun2023berlin,doering2022pqc5g,OpenQKD2024-D75} & 2020--2023 & Star-shaped dark-fibre metro network; three QKD nodes at the central site with looped links to remote sites & DV (ID Quantique and Toshiba systems) plus an Open Quantum Safe PQC key-exchange system in the quantum layer & Yes: PQC key exchange as an additional key source; PQC KEM/signature-secured KMS; keys imported into an HSM and exposed over ETSI 014 to L1 and L3 encryptors & In our corpus, the only provider testbed with a PQC-secured key-management plane and hybrid key delivery to commercial encryptors \\
MadQCI, Madrid (UPM/Telef\'onica)~\cite{martin2024madqci,buruaga2025tls} & 2019-- & 28 QKD modules from several vendors in a multi-tenant production telecom network; SDN-controlled & DV and CV, multi-vendor & Yes: hybrid TLS in SDN~\cite{buruaga2025tls}; VMuckle deployment planned by its authors~\cite{Buruaga2025-VMuckle} & Disaggregated SDN-QKD architecture in a production network \\
Eindhoven testbed~\cite{garcia2023camad,jsac2025_hybrid} & 2023--2025 & Laboratory/metro link & DV & Yes: three-way hybrid in TLS, IPsec, MACsec & Cross-layer overhead measurements \\
Toshiba hybrid cryptosystem (Muckle++)~\cite{garms2024experimental} & 2024 & Point-to-point QKD hardware & DV (Toshiba) & Yes: QKD and PQ KEM keys combined; PQ signatures & Experimental integration of the two key sources on commercial QKD hardware, with a HAKE-based proof (Table~\ref{tab:proofs}) \\
Vigo--Madrid prototypes~\cite{blanco2025qkdkem,hybrid_ipsec2025} & 2025 & Two ID Quantique Cerberis XGR nodes over dedicated dark fibre in Vigo; TLS/IPsec endpoints in Madrid & DV (commercial) & Yes: QKD-KEM in TLS (OpenSSL provider) and IKEv2 (strongSwan) & Handshake and IKE measurements with remote key-management entities (Table~\ref{tab:protocol_overview}) \\
\bottomrule
\end{tabular}
\end{table*}

Across these networks, four findings are consistent and sourced. (1)~Multi-vendor interoperability at the key-delivery interface (ETSI 014/004) is demonstrated~\cite{OpenQKD2024-D75}. (2)~Trusted-node relaying is the range-extension mechanism in each multi-span network in Table~\ref{tab:testbeds}~\cite{chen2021integrated,cao2022evolution}. (3)~Where PQC has been used, it has been used first to secure the QKD network's own control and key-management plane~\cite{yang2021pqcauth,geitz2023hybrid}, and only more recently as a second key source in TLS, IPsec, and MACsec~\cite{jsac2025_hybrid,Buruaga2025-VMuckle}. (4)~In the TLS measurements that report handshake latency~\cite{garcia2023camad,blanco2025qkdkem}, the added latency is attributed to key retrieval rather than to PQC arithmetic. These findings concern different metrics, which Table~\ref{tab:measurements} keeps apart: application throughput, QKD secret-key rate, handshake latency, and rekey interval are not interchangeable, and a high application throughput (for example, the 80~Gb/s tunnel of Alia et al.) says nothing about the secret-key rate that limits how often keys can be refreshed (7.4~kb/s in the same trial)~\cite{alia2024ipsec}. What the evidence does not yet support is a claim about carrier-grade availability, long-term operational cost, or security incidents, because we found no such data in peer-reviewed form for the networks in Tables~\ref{tab:testbeds} and~\ref{tab:testbeds_hybrid}.

\begin{table*}[t]
\caption{Reported measurements by metric type, with the experimental conditions stated by each source. Values are not comparable across sources. n.r.\ = not reported in the cited source.}
\label{tab:measurements}
\scriptsize\setlength{\tabcolsep}{3.5pt}
\begin{tabular}{@{}L{2.6cm}L{2.2cm}L{3.8cm}L{5.6cm}@{}}
\toprule
\textbf{Source} & \textbf{Metric} & \textbf{Reported value} & \textbf{Experimental conditions (as reported)} \\
\midrule
Rubio Garc\'ia et al.~\cite{garcia2023camad} & Handshake latency & About $+68\%$ total handshake time (triple hybrid); about $+104\%$ with an emulated fast ETSI 014 API & Triple-hybrid TLS 1.3 in an experimental network scenario~\cite{garcia2023camad} \\
Blanco-Romero et al.~\cite{blanco2025qkdkem} & Handshake latency & About 100 ms per ETSI 014 call; full handshake 300--350 ms vs.\ 13--18 ms for pure PQC & Two ID Quantique Cerberis XGR nodes on dedicated dark fibre in Vigo; TLS endpoints in Madrid; keys carried over HTTPS between key-management entities and endpoints \\
Rubio Garc\'ia et al.~\cite{jsac2025_hybrid} & Handshake performance; packet overhead & ``Non-substantial'' decrease; negligible added packet overhead (qualitative) & Qualitative result only; used here as reported~\cite{jsac2025_hybrid} \\
Blanco-Romero et al.~\cite{hybrid_ipsec2025} & IKE handshake latency; bandwidth overhead & Qualitative: parallel QKD-KEM composition avoids the per-exchange latency of sequential RFC 9370 exchanges; minimal bandwidth overhead in identifier-based mode & strongSwan; ID Quantique hardware; ETSI 004 and 014 interfaces \\
Alia et al.~\cite{alia2024ipsec} (no PQC) & Application throughput & 80 Gb/s (one tunnel); 99.62 Gb/s aggregate (12 tunnels at 8.39 Gb/s each) & 46 km deployed fibre, Singapore; two data centres; air-gapped; ID Quantique Clavis XGR (decoy-state BB84); FortiGate firewalls; AES-256-GCM; keys via ETSI 014; 24 h stress test within 45 days of operation \\
Alia et al.~\cite{alia2024ipsec} (no PQC) & QKD secret-key rate & 7.4 kb/s average (QBER 0.8\%) & As above \\
Alia et al.~\cite{alia2024ipsec} (no PQC) & Rekey interval & 120 s (0.03\% of generated keys consumed, single tunnel; 0.3\%, 12 tunnels) & As above; no loss of tunnel connectivity or performance reported \\
Jinan network~\cite{yang2021pqcauth,wang2021npjqi} & Continuous operation & 36 days & 14 user nodes and 5 optical switching nodes; PQ signatures (Aigis-Sig) authenticate the QKD classical channel \\
Rubio Garc\'ia et al.~\cite{rubio2024comcom} & Key-establishment time & About $+117\%$ for hybrid PQC--QKD TLS (concatenation and XOR variants); PQC-only about 9\% faster than classical & Hybrid TLS~1.2 integration with commercial QKD devices~\cite{gsma2024ig18}, Eindhoven QKD testbed \\
Muckle+~\cite{bruckner2023mucklePlus} & Handshake latency & 0.4--1.6 s per stage for most KEM/signature choices, of which about 0.3 s is QKD key retrieval; longer in the multi-path setting & Python prototype; small field QKD network (3 links, 2 disjoint paths); ETSI 014; Windows notebook (Intel i5, 8 GB) \\
VMuckle~\cite{Buruaga2025-VMuckle} & Handshake CPU and bandwidth & 1.5--2.2 G cycles; 42.2 KB for ML-KEM-1024 + ML-DSA-87 + QKD + ECDH vs.\ 17.29 KB for EAP-TLS & Python prototype with liboqs; simulated QKD; Ubuntu notebook (Intel i7-8665U, 16 GB) \\
Battarbee et al.~\cite{battarbee2025kemauth} & Handshake CPU and bandwidth & 1.6--6.0 G cycles; 29.2--153.9 KB (e.g.\ ML-KEM-768: 1.7 G cycles, 31.3 KB) & Python prototype with liboqs; simulated QKD; 5 runs; Windows notebook (Intel i5, 8 GB) \\
Clermont \& Henrich~\cite{clermont2026best} & Hybridisation latency & Median 11--13 ms with ML-KEM; about $+100$ ms when 20 ms per-message latency is added; up to about 150 ms with Classic McEliece level 5 & Go implementation in Docker; simulated KMS with pre-established QKD keys; 1000 runs per configuration; 8 vCPU server \\
\bottomrule
\end{tabular}
\end{table*}
\section{Standards and Policy Landscape}\label{sec:standards}

Table~\ref{tab:standards} collects the documents that govern hybrid PQC--QKD deployment. Two features of this landscape are frequently misstated in the technical literature and deserve emphasis.

First, the standards for combining keys exist and are mature: NIST SP~800-56C~Rev.\,2 permits a hybrid shared secret~\cite{nist2020sp80056c}, SP~800-227 gives KEM guidance~\cite{nist2025sp800227}, ETSI TS~103~744 specifies hybrid key exchanges~\cite{etsi_ts103744}, the IETF has three RFCs for IKEv2~\cite{rfc8784,rfc9370,rfc9867} and two RFCs for classical+PQC hybrid key exchange in TLS~1.3~\cite{rfc9954,rfc10024}, and ITU-T X.1714 covers combination inside QKD networks~\cite{itut_x1714}. A design that introduces a new combiner should therefore state why the existing constructions do not meet its goal; information-theoretic key secrecy, which the KDF-based constructions do not provide, is one such reason (Section~\ref{sec:combiner_constructions}).

Second, the agencies that set procurement policy in the United States, the United Kingdom, France, Germany, the Netherlands, and Sweden have stated that QKD is not an approved or recommended replacement for PQC, citing the need for special-purpose hardware, the authentication problem, the trusted-node problem, the gap between proofs and devices, and the absence of a certification path~\cite{nsa2020qkd,ncsc2020quantum,agencies2024qkdposition}; CNSA~2.0 mandates PQC algorithms and timelines without QKD~\cite{nsa2022cnsa2}, as do NSM-10~\cite{whitehouse2022nsm10}, the CISA/NSA/NIST readiness guidance~\cite{cisa2023quantum}, NIST IR~8547~\cite{nist2024ir8547}, and the EU recommendation~\cite{eu2024recommendation}. The European position paper explicitly allows QKD ``in addition to'' PQC, which is the hybrid position; the QKD community has responded to the agencies' objections~\cite{renner2023debate,stebila2010case}, and a recent critical analysis from a group that builds hybrid systems examines which deployed QKD use cases actually gain over PQC~\cite{aquina2025critical}. A survey of hybrid systems must nonetheless state plainly that the policy baseline is PQC, that QKD is an optional additional layer in the eyes of these agencies, and that the burden of evidence for adding it is on the deployment. The certification frameworks that could discharge that burden (ETSI GS QKD 016 protection profile~\cite{etsi_qkd_016}, ISO/IEC 23837~\cite{iso23837}) are recent and, to our knowledge, no hybrid PQC--QKD product has been evaluated against them together with FIPS~140-3 validation of its PQC component.

\begin{table*}[t]
\caption{Standards and policy documents governing hybrid PQC--QKD deployment.}
\label{tab:standards}
\footnotesize\setlength{\tabcolsep}{4pt}
\begin{tabular}{@{}L{1.4cm}L{3.4cm}L{1.9cm}L{7.4cm}@{}}
\toprule
\textbf{Body} & \textbf{Document} & \textbf{Status} & \textbf{Relevance to hybrid PQC--QKD} \\
\midrule
NIST & FIPS 203/204/205~\cite{NIST-FIPS203,NIST-FIPS204,NIST-FIPS205} & Final (2024) & PQC algorithms; draft FIPS 206 and HQC pending~\cite{nist2025hqc} \\
NIST & SP 800-56C Rev.\,2~\cite{nist2020sp80056c} & Final (2020) & Hybrid shared secret $Z\|T$ in approved key derivation \\
NIST & SP 800-227~\cite{nist2025sp800227} & Final (2025) & KEM usage and combination guidance \\
NIST & IR 8547~\cite{nist2024ir8547} & Draft (2024) & Deprecation timeline for quantum-vulnerable algorithms (2030/2035) \\
IETF & RFC 8784~\cite{rfc8784} & Standard (2020) & PPK (incl.\ QKD key) mixing in IKEv2 \\
IETF & RFC 9242, RFC 9370~\cite{rfc9242,rfc9370} & Standard (2022, 2023) & IKE\_INTERMEDIATE; additional (PQ) key exchanges in IKEv2 \\
IETF & RFC 9867~\cite{rfc9867} & Standard (2025) & PPK in IKE\_INTERMEDIATE and CREATE\_CHILD\_SA; refreshed QKD keys \\
IETF & RFC 9954; RFC 10024~\cite{rfc9954,rfc10024} & Published (2026) & Classical+PQC hybrid key exchange in TLS 1.3; template for QKD extensions; no QKD group registered \\
ETSI & TS 103 744~\cite{etsi_ts103744} & Published (2020) & Quantum-safe hybrid key exchange constructions (concatenation, cascade) \\
ETSI & GS QKD 004, 014~\cite{etsi_qkd_004,etsi_qkd_014} & Published & Key-delivery APIs used by the surveyed protocol integrations \\
ETSI & GS QKD 015, 018~\cite{etsi_qkd_015,etsi_qkd_018} & Published (2022) & SDN control and orchestration of QKD networks \\
ETSI & GS QKD 020~\cite{etsi_qkd_020} & Published (2026) & REST-based interoperable key-management-system API (inter-KMS key relay) \\
ETSI & GS QKD 016~\cite{etsi_qkd_016} & Published (V1.1.1 2023; V2.1.1 2024) & Common Criteria protection profile for prepare-and-measure QKD modules; first commercial module certified against it by BSI \\
ITU-T & Y.3800 series~\cite{itut_y3800} & Published & QKD network architecture, key management, control \\
ITU-T & X.1710, X.1714~\cite{itut_x1710,itut_x1714} & Published & QKDN security framework; key combination and confidential key supply \\
ISO/IEC & 23837-1/-2~\cite{iso23837} & Published (2023) & Security requirements and evaluation methods for QKD modules \\
IEEE & 802.1AE~\cite{ieee8021ae}; 802.1X (MKA) & Standard & MACsec; VMuckle supplies the MKA root key without modifying these standards~\cite{Buruaga2025-VMuckle} \\
ETSI & TS 104 146 (VMuckle-ES)~\cite{abelard2026assessment} & Draft (V0.0.4, 2026) & Authenticated quantum-safe hybrid key establishment including QKD key identifiers, as reported in~\cite{abelard2026assessment} \\
GSMA & IG.18~\cite{gsma2024ig18} & Whitepaper (2024) & Industry analysis of hybrid QKD--PQC scenarios, standards activity and gaps, proofs of concept \\
NSA & QKD statement~\cite{nsa2020qkd}; CNSA 2.0~\cite{nsa2022cnsa2} & Policy & QKD not recommended for NSS; PQC algorithm mandates and dates \\
NCSC (UK) & Quantum security technologies~\cite{ncsc2020quantum} & Policy & Does not endorse QKD for government/military use \\
ANSSI, BSI, NLNCSA, SNCSA & Joint QKD position paper~\cite{agencies2024qkdposition} & Policy (2024) & QKD not a substitute for PQC; acceptable only in addition to PQC \\
White House; CISA/NSA/NIST; EU & NSM-10~\cite{whitehouse2022nsm10}; readiness factsheet~\cite{cisa2023quantum}; Recommendation 2024/1101~\cite{eu2024recommendation} & Policy & Migration mandates and timelines; PQC-centred \\
\bottomrule
\end{tabular}
\end{table*}

\section{Deployment Considerations}\label{sec:deployment}

\subsection{Reference architecture}

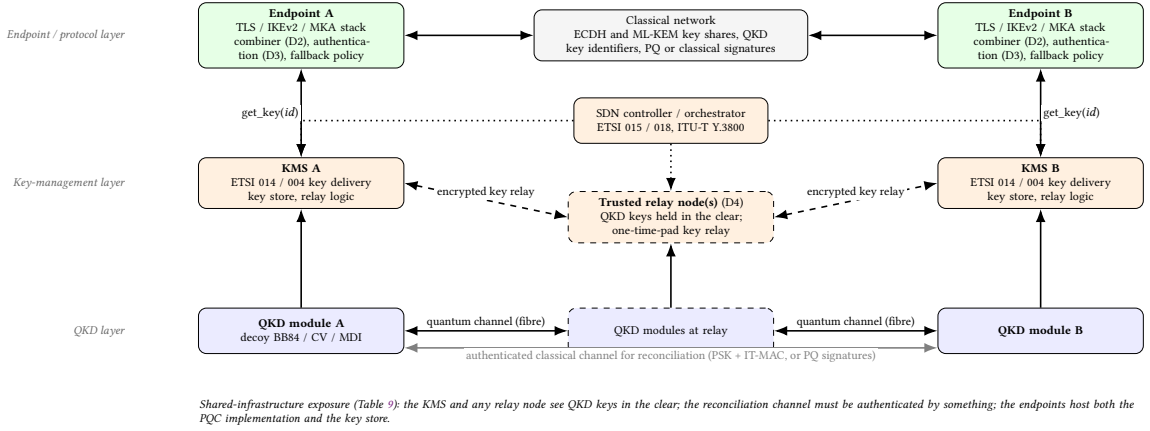
\begin{figure*}[t]
\centering
\resizebox{\textwidth}{!}{%
\begin{tikzpicture}[
  font=\scriptsize,
  >=Latex,
  box/.style={draw, rounded corners, align=center, minimum height=0.8cm, fill=black!4, text width=3.4cm},
  q/.style={box, fill=blue!8},
  k/.style={box, fill=orange!12},
  e/.style={box, fill=green!10},
  lbl/.style={font=\scriptsize, fill=white, inner sep=1pt},
  layer/.style={font=\scriptsize\itshape, anchor=east, text=black!60}
]
\node[layer] at (-3.0,0)    {Endpoint / protocol layer};
\node[layer] at (-3.0,-2.6) {Key-management layer};
\node[layer] at (-3.0,-5.2) {QKD layer};
\node[e] (A) at (0,0) {\textbf{Endpoint A}\\TLS / IKEv2 / MKA stack\\combiner (D2), authentication (D3), fallback policy};
\node[e] (B) at (13.0,0) {\textbf{Endpoint B}\\TLS / IKEv2 / MKA stack\\combiner (D2), authentication (D3), fallback policy};
\node[box, text width=4.6cm] (net) at (6.5,0) {Classical network\\ECDH and ML-KEM key shares, QKD key identifiers, PQ or classical signatures};
\draw[<->, thick] (A) -- (net);
\draw[<->, thick] (net) -- (B);
\node[k, text width=3.2cm] (orch) at (6.5,-1.5) {SDN controller / orchestrator\\ETSI 015 / 018, ITU-T Y.3800};
\node[k] (KA) at (0,-2.6) {\textbf{KMS A}\\ETSI 014 / 004 key delivery\\key store, relay logic};
\node[k] (KB) at (13.0,-2.6) {\textbf{KMS B}\\ETSI 014 / 004 key delivery\\key store, relay logic};
\node[k, dashed, text width=3.4cm] (TN) at (6.5,-3.2) {\textbf{Trusted relay node(s)} (D4)\\QKD keys held in the clear; one-time-pad key relay};
\draw[<->, thick] (A) -- node[lbl, left]{get\_key($\mathit{id}$)} (KA);
\draw[<->, thick] (B) -- node[lbl, right]{get\_key($\mathit{id}$)} (KB);
\draw[<->, thick, dashed] (KA.east) -- node[lbl, above]{encrypted key relay} (TN.west);
\draw[<->, thick, dashed] (TN.east) -- node[lbl, above]{encrypted key relay} (KB.west);
\draw[->, thick, dotted] (orch.west) -| (KA.north);
\draw[->, thick, dotted] (orch.east) -| (KB.north);
\draw[->, thick, dotted] (orch.south) -- (TN.north);
\node[q] (QA) at (0,-5.2) {\textbf{QKD module A}\\decoy BB84 / CV / MDI};
\node[q] (QB) at (13.0,-5.2) {\textbf{QKD module B}};
\node[q, dashed] (QT) at (6.5,-5.2) {QKD modules at relay};
\draw[<->, thick] (QA.east) -- node[lbl, above]{quantum channel (fibre)} (QT.west);
\draw[<->, thick] (QT.east) -- node[lbl, above]{quantum channel (fibre)} (QB.west);
\draw[<->, thick, gray] ([yshift=-0.3cm]QA.east) -- node[lbl, below]{authenticated classical channel for reconciliation (PSK + IT-MAC, or PQ signatures)} ([yshift=-0.3cm]QB.west);
\draw[->, thick] (QA) -- (KA); \draw[->, thick] (QB) -- (KB); \draw[->, thick] (QT) -- (TN);
\node[anchor=north west, text width=16.5cm, align=left, font=\scriptsize\itshape] at (-1.9,-6.3) {Shared-infrastructure exposure (Table~\ref{tab:threats}): the KMS and any relay node see QKD keys in the clear; the reconciliation channel must be authenticated by something; the endpoints host both the PQC implementation and the key store.};
\end{tikzpicture}}
\caption{Reference architecture of a hybrid PQC--QKD deployment, annotated with the taxonomy dimensions at which each design choice is made. The surveyed integrations follow this pattern; designs differ in whether the combiner runs at the endpoint or inside the KMS, in the authentication source, and in whether relay nodes are present.}
\Description{A layered block diagram. At the top, two endpoints exchange PQC key shares and QKD key identifiers over a classical network. Below them, key management systems at each side deliver QKD keys over ETSI 014 or 004 interfaces and relay keys through a dashed trusted relay node, coordinated by an SDN orchestrator. At the bottom, QKD modules are connected by quantum channels through the relay and by an authenticated classical channel.}
\label{fig:architecture}
\end{figure*}

Figure~\ref{fig:architecture} shows the components that the surveyed deployments share, and where the design choices of the taxonomy are made. The QKD layer produces keys per span; a key-management system (KMS) relays, stores, and serves keys through ETSI~014/004 and, in networked deployments, is orchestrated through ETSI~015/018 or ITU-T Y.3800-series interfaces; endpoints request keys by identifier; and the combiner runs either at the endpoint (Muckle family, TLS and IKEv2 designs) or inside the KMS (X.1714 pattern). The trusted-node and KMS boxes are where the shared-infrastructure threats of Table~\ref{tab:threats} concentrate.

\subsection{Topology and trust}

QKD range per span is bounded by fibre loss, so any network larger than one span relays keys through nodes that decrypt and re-encrypt them; each deployment in Table~\ref{tab:testbeds} that spans more than one link does this~\cite{mehic2020quantum,cao2022evolution,chen2021integrated}. MDI- and TF-QKD relocate trust away from the relay's measurement but still require a chain of end nodes over long distances. For a hybrid deployment, the practical implication is that the PQC component should be terminated end-to-end so that a relay compromise exposes only the QKD input, and that the relay itself should be protected with PQC-authenticated and PQC-encrypted key transport~\cite{geitz2023hybrid}. Topology choice (point-to-point, ring, hierarchical with a backbone) follows from where trusted nodes can be physically secured, not from key-management complexity; Mehic et al.\ and Cao et al.\ treat routing, key pools, and scalability in detail~\cite{mehic2020quantum,cao2022evolution}.

\subsection{Key management and operations}

Hybrid deployments manage two key sources with different lifecycles: QKD keys are generated continuously and buffered in per-link key pools, whereas PQC keys are generated per session. Operational parameters (pool sizes, refresh intervals, QBER alarm thresholds, availability targets) are set by operators and vendors and, in the sources we reviewed, are reported only in project deliverables~\cite{OpenQKD2024-D75}; we do not reproduce specific values here because no peer-reviewed source states them as general recommendations. What the standards do fix is the interface: ETSI~014 key requests carry key identifiers and sizes, ETSI~004 supports stateful sessions, and RFC~9867 defines how a refreshed PPK is taken into use in IKEv2 without re-establishing the SA. Fallback behaviour when a QKD key is unavailable is a policy decision that must be explicit and negotiated (Table~\ref{tab:threats}, last row); VMuckle's configuration modes and RFC~8784's optional-PPK flag are the published mechanisms we found.

\subsection{Cost}

QKD deployment cost is dominated by transceiver hardware and by the fibre and secured sites needed for trusted nodes; the PQC component is software and adds little~\cite{Fedorov2023-Deploying-Hybrid}. Mehic et al.~\cite{mehic2020quantum} discuss QKD network cost structure, and Jia et al.~\cite{jia2023cost} optimise key-pool provisioning cost; beyond these, we found no peer-reviewed cost or benefit--cost analysis of a hybrid PQC--QKD deployment, and we regard the absence of such an analysis as an open problem (Section~\ref{sec:open}) rather than attempt one here.

\section{Open Problems}\label{sec:open}

The preceding analysis yields the following open problems, each stated with the evidence that it is open and with concrete research questions.

\paragraph{O1. Authentication of the QKD classical channel in hybrid systems.} The proofs we reviewed assume an authenticated QKD classical channel; deployments bootstrap it with pre-shared keys or with PQ signatures~\cite{alleaume2014using,wang2021npjqi,yang2021pqcauth,geitz2023hybrid}. Muckle and Muckle+ sit on either side of the PSK-versus-signature choice, and Battarbee et al.~\cite{battarbee2025kemauth} and Clermont and Henrich~\cite{clermont2026best} have now shown that KEM-based authentication can replace both, the latter with a conditional information-theoretic guarantee. \emph{Remaining questions:} What exactly is preserved when the authenticating KEM is broken after the exchange? Can the QKD post-processing transcript or session identifier be bound into the hybrid AKE when key identifiers are unavailable, as H\"ovelmanns et al.\ suggest~\cite{hoevelmanns2025qkdoracles}, so that the QKD phase is tied to the session by the same mechanism? How do these designs interact with the PKI that a QKD network operator must run?

\paragraph{O2. Proofs that include the full key-delivery path.} H\"ovelmanns et al.~\cite{hoevelmanns2025qkdoracles} model an ETSI 014-like interface with key identifiers inside the CK+ model and show that unbound identifiers enable dependent-key attacks; Mosca et al.~\cite{mosca2013qkdake} and Ferreira~\cite{ake2026_pqc_qkd} model the QKD phase, and the remaining proofs treat the QKD key as an abstract source (Table~\ref{tab:proofs}). \emph{Questions:} Extend interface models to the stateful ETSI~004 interface, multi-tenant KMSs, and inter-KMS relay over ETSI GS QKD 020; model $\varepsilon$-secure rather than perfectly uniform QKD keys, which H\"ovelmanns et al.\ leave open and for which Ferreira finds that deployed parameters give non-meaningful bounds~\cite{ake2026_pqc_qkd}; give a composable statement that connects a QKD protocol's $\varepsilon_{\mathrm{QKD}}$ to the AKE advantage through the KMS.

\paragraph{O3. Shared-infrastructure threat models.} The proofs we reviewed cover the rows of Table~\ref{tab:threats} at most partially. H\"ovelmanns et al.'s QKD-OVERRIDE oracle captures an attacker-set QKD key on one end-to-end link~\cite{hoevelmanns2025qkdoracles}. \emph{Questions:} Define corruption models for KMS and trusted-node compromise that fall between ``QKD ideal'' and ``QKD fully broken'', and relate per-key corruption to the compromise of a relay or KMS that serves many sessions; determine which combiners keep the PQC guarantee in that setting.

\paragraph{O4. Certification versus proof.} Device-level certification of QKD modules (ETSI GS QKD 016, ISO/IEC 23837) and FIPS~140-3 validation of PQC modules are separate processes with no joint profile. \emph{Questions:} What would a protection profile for a hybrid endpoint contain? Can the assumptions of the protocol proof be mapped onto evaluation assurance requirements so that certification discharges them?

\paragraph{O5. Formal verification of hybrid handshakes.} The classical+PQC TLS hybrid has been analysed with the tools used for TLS~1.3; we found no Tamarin or ProVerif model of a QKD-extended TLS or IKEv2 (RFC~9867) handshake, or of the VMuckle exchange and its hand-off to MKA. \emph{Questions:} Model the QKD key-ID exchange and PPK-ID exchange symbolically; check downgrade and identifier-confusion properties.

\paragraph{O6. Standardised bindings.} To our knowledge, no IANA-registered TLS group carries a QKD key identifier, and we found no profile that binds RFC~9867 PPK identifiers or ETSI~014 key identifiers across vendors. Such a binding must also authenticate the identifier as part of the transcript~\cite{hoevelmanns2025qkdoracles}. \emph{Questions:} What is the minimal interoperable binding, and can it be specified without freezing a particular KMS design?

\paragraph{O7. Deployment economics and operational evidence.} We found no peer-reviewed availability, incident, or cost data for a hybrid PQC--QKD deployment; among the sources we reviewed, project deliverables~\cite{OpenQKD2024-D75} are the only ones that report such data. \emph{Questions:} Which metrics should a hybrid deployment report so that its security and cost claims can be compared with the classical+PQC baseline?

\paragraph{O8. The value question.} Given that the combined key is consumed by a computational cipher and that agencies regard QKD as an optional additional layer~\cite{agencies2024qkdposition}, the argument for adding QKD reduces to hedging against a break of lattice assumptions within the confidentiality lifetime of the data; Aquina et al.~\cite{aquina2025critical} examine deployed use cases from this standpoint and Renner and Wolf~\cite{renner2023debate} argue the other side. \emph{Questions:} For which data lifetimes and threat models does QKD's hedge exceed that of algorithm diversity within PQC (e.g., ML-KEM plus HQC, as NIST's selection of a code-based KEM intends), and at what cost?

\section{Conclusion}\label{sec:conclusion}

Hybrid PQC, QKD systems have relatively little published work on them and a larger corpus of related work. The larger corpus on hybrid classical and PQC key exchange includes deployed and standardized systems and protocols and hybrid key exchange combiners, such as dual-PRF combiners used in TLS and other deployed systems documented in RFCs and used in IKEv2. In the relatively small corpus on hybrid PQC, QKD systems, systems are described in which a QKD key is added as an input to a key exchange protocol. Our review highlights differences between the relatively small corpus, including works by Muckle et al., Battarbee et al., Ferreira, and Clermont and Henrich, and H\"ovelmanns et al., and deployed hybrid PQC--QKD systems, including TLS, IPsec, and MACsec implementations using keys provided per ETSI standards, with exception of Muckle+, Muckle++, and an IPsec implementation by Garms et al., all of which have security proofs and include QKD hardware; widespread use of PQC within the QKD network rather than for hybrid key exchange in test networks described as hybrid; lack of use of PQC in at least one widely cited IPsec implementation with QKD security; and greater importance of binding exchanged and QKD keys and QKD key identifiers to a session than using a KDF or XOR combiner. KDF-based hybrid combiners in standards documents cannot provide information-theoretic secrecy for hybrid keys, though this does not impact confidentiality of upper-layer messages when using a computationally secure AEAD, and information-theoretic key secrecy is possible in some hybrid systems when using a secure QKD system, as demonstrated by H\"ovelmanns et al. (using an XOR combiner and information-theoretic MACs) and Clermont and Henrich (using a multi-input KDF modeled as a random oracle).

What is not well understood are the relationships of the protocol model to possible system implementations, including approaches to authenticating QKD and other key management protocols, alternative end-user protocol models to the oracle model used in ETSI 014, such as stateful, multitenant, relayed key delivery, and $\epsilon$-secure keys; shared trusted infrastructure and key management; device certification approaches; and formal methods for verifying end-user handshake protocols. The most important survey finding from a policy standpoint is a nontechnical one, procuring agencies see PQC as a necessity and QKD as an option that must be justified on the basis of advantages of hybrid QKD and PQC systems for certain data storage durations and threat models. There are no peer-reviewed publications providing evidence on the operation and economics of hybrid QKD and PQC systems and conditions for their use. Research on these topics would be a valuable contribution by others in the research community.

\bibliographystyle{plain}
\bibliography{ref_final}

\end{document}